\pdfoutput=1
\documentclass[useAMS,usenatbib]{mnras}

\usepackage{graphicx}
\usepackage{bigstrut}
\usepackage{enumitem}
\usepackage{hyperref}
\newcommand{\IleeTime}{389}
\newcommand{\ORPTime}{440}
\newcommand{\FinalTime}{2043}

\title[GIs in a protosolar-like disc I]{Gravitational instabilities in a protosolar-like disc I: dynamics and chemistry}
\author[M.\ G.\ Evans et al.]{M.\ G.\ Evans$^{1}$\thanks{E-mail: py09mge@leeds.ac.uk},
J.\ D.\ Ilee$^{2}$, A.\ C.\ Boley$^{3}$, P.\ Caselli$^{4}$, R.\ H.\ Durisen$^{5}$, \and T.\ W.\ Hartquist$^{1}$ and J.\ M.\ C.\ Rawlings$^{6}$\smallskip\\ 
$^{1}$School of Physics \& Astronomy, University of Leeds, Leeds LS2 9LN, UK\\
$^{2}$SUPA, School of Physics \& Astronomy, University of St Andrews, North Haugh, St Andrews, Scotland, KY16 9SS, UK\\
$^{3}$Department of Physics \& Astronomy, 6224 Agricultural Road, Vancouver, BC V6T 1Z1, Canada\\
$^{4}$Max-Planck-Institut f$\ddot{u}$r extraterrestrische Physik, Gie$\beta$enbachstra$\beta$e, 85741 Garching bei M$\ddot{u}$nchen, Germany\\
$^{5}$Department of Astronomy, Indiana University, 727 East 3rd Street, Swain West 319, Bloomington, IN 47405, USA\\
$^{6}$Department of Physics \& Astronomy, University College London, London WC1E 6BT, UK}
\begin{document}

\date{Accepted 2015 July 24.  Received 2015 July 22; in original form 2015 February 20}

\maketitle

\label{firstpage}

\begin{abstract}
To date, most simulations of the chemistry in protoplanetary discs have used 1+1D or 2D axisymmetric $\alpha$-disc models to determine chemical compositions within young systems. This assumption is inappropriate for non-axisymmetric, gravitationally unstable discs, which may be a significant stage in early protoplanetary disc evolution.  Using 3D radiative hydrodynamics, we have modelled the physical and chemical evolution of a $0.17\,\mathrm{M}_{\odot}$ self-gravitating disc over a period of 2000\,yr.  The $0.8\,\mathrm{M}_{\odot}$ central protostar is likely to evolve into a solar-like star, and hence this Class 0 or early Class I young stellar object may be analogous to our early Solar System.  Shocks driven by gravitational instabilities enhance the desorption rates, which dominate the changes in gas-phase fractional abundances for most species.  We find that at the end of the simulation, a number of species distinctly trace the spiral structure of our relatively low-mass disc, particularly CN.  We compare our simulation to that of a more massive disc, and conclude that mass differences between gravitationally unstable discs may not have a strong impact on the chemical composition.  We find that over the duration of our simulation, successive shock heating has a permanent effect on the abundances of HNO, CN and NH$_3$, which may have significant implications for both simulations and observations. We also find that HCO$^+$ may be a useful tracer of disc mass.  We conclude that gravitational instabilities induced in lower mass discs can significantly, and permanently, affect the chemical evolution, and that observations with high-resolution instruments such as ALMA offer a promising means of characterising gravitational instabilities in protosolar discs.

\end{abstract}


\section{Introduction}

In the very early stages of stellar formation, the surrounding protoplanetary disc can contain a mass comparable to that of the central protostar \citep[e.g.][]{Machida2011}. As a result, self-gravity within the disc is significant and the disc can become unstable if gravitational forces overcome pressure and shear forces; a criterion characterised by the Toomre parameter \citep{Toomre1964}. Gravitational instabilities (GIs) formed in the disc via this process produce spiral density waves that efficiently drive angular momentum transport outwards and mass accretion inwards. As these spiral waves grow, they can produce shocks that heat the disc material locally \citep[e.g.][]{Harker2002, Boley2008, Bae2014}. This episodic heating can enhance the rates of chemical reactions, such as desorption of volatiles and reactions with high activation energies, which may significantly affect the chemical composition of discs.

\smallskip

Many authors have investigated the chemical properties and evolution of protoplanetary discs \citep[see][Table 3 for an extensive review]{Henning2013}. However, most simulations have featured a passive or weakly viscous $\alpha$-disc with 2D axisymmetric structure and local mass transport. Hence, this prescription is likely inappropriate for early phases of disc evolution if GIs are present driving non-axisymmetric structure. To date, only one study has explored the effect of GIs and subsequent shock-heating on disc chemistry.

\smallskip

\citet{Ilee2011}, hereafter I2011, simulated a massive protoplanetary disc throughout an evolutionary phase that may be applicable to an FU Orionis outburst. The system represents a Class 0/early Class I object and the relatively massive ($0.39\,\mathrm{M}_{\odot}$) disc is appropriate for a protostar that will likely become an F star ($\approx1.4\,\mathrm{M}_{\odot}$) at the end of the accretion phase. The authors found that shock-heating within the disc induced temperature peaks that desorbed various chemical species from the surface of dust grains and enhanced the rates of reactions that were otherwise not energetically favourable, therefore increasing some gas-phase abundances. In contrast to simulations of more evolved, less dynamic systems, they found that the midplane is the hottest region and hence rich in chemistry. All species that they consider are tracers of the spiral shocks, and \citet{Douglas2013} showed that, encouragingly, these gas-phase tracers of disc dynamics should be detectable using ALMA at a distance of 100\,pc. Furthermore, \citet{Dipierro2014} have recently shown that ALMA can readily detect spiral structure in continuum emission, within Class 0/I systems located at distances comparable to TW Hydrae, Taurus-Auriga and Ophiucus star-forming regions. Hence this will allow the empirical assessment of whether young protoplanetary discs around intermediate mass stars are indeed gravitationally unstable. 

\smallskip

Simulations of gravitationally unstable discs can be used to determine which observable features are the most promising for detection, should the instabilities be present. Observations can then test and constrain models for the important dynamical and chemical evolution of young protoplanetary discs. If gravitational instabilities are not detected in real objects, then the model parameters can be refined until GIs can be satisfactorily excluded as a possibility in young systems, or the simulations can be improved to identify actual observable signatures. The motivation for this paper is to provide a first step in this direction for a protosolar disc, i.e.\ a young disc surrounding a protostar that will become a Solar-like star at the end of the accretion phase.

\smallskip

In Section \ref{sec:Models} we outline the physical and chemical models we have used and how they differ from those used in the simulation of the more massive disc.  Section \ref{sec:Results} contains results for time-dependent fractional gas-phase abundances determined from individual fluid elements and column density maps of the entire disc in order to assess the differences in composition between the more massive disc, the lower mass disc at different times, and discs featured in other published works. Finally, in Section \ref{sec:Conclusion} we present conclusions based on our results and discuss future research.

\section[Models]{Disc Models}
\label{sec:Models}
\subsection{Dynamical simulation}

A radiation hydrodynamics simulation of a $0.17\,\mathrm{M}_{\odot}$ protoplanetary disc is run over a period of approximately 2000\,yr and used as the input into our chemical model. The disc surrounds a central protostar of $0.8\,\mathrm{M}_{\odot}$, which we envisage will evolve into a Solar-like star. The majority of the mass in the disc initially ranges between r $\approx$ 6--41\,au from the central protostar, but spreads partially during its evolution as a consequence of angular momentum transport and accretion, resulting in a disc that spans r $\approx$ 5--54\,au at the end of evolution. The system represents a Class 0 or early Class I object that can potentially be compared to the early Solar System.

\smallskip

The physical simulation is performed using {\sc{CHYMERA}} \citep{Boley2007}, which solves the equations of hydrodynamics on a regularly spaced cylindrical grid. Outflow boundaries are used at the inner, outer, and top grid edges and mirror symmetry is assumed along the midplane of the disc. Self-gravity is calculated directly through cyclic reduction, which  is a numerical method for solving large linear systems by repeatedly splitting the problem, with the boundary potential determined by a spherical harmonics expansion. An indirect potential is used for capturing the star-disc interactions \citep[e.g.][]{Michael2010} and the \citet{Boley2007a} equation of state is used for a fixed H$_2$ ortho-para ratio of 3:1 in a Solar mixture gas. We use the \citet{Boley2007b} radiative transfer algorithm, which combines flux-limited diffusion with ray tracing in the vertical direction. The radiative routine calculates a separate radiative time step, and subcycles in the event that the hydrodynamics time step becomes much larger than the radiative step. A maximum iteration of 8 sub cycles is used. If this limit is reached, then one last step is used to synchronize the hydrodynamics and radiative times. To ensure stability, energy is not allowed to change in any one cell by more than 10 per cent \citep[see][]{Boley2006}; such limiters are typically only necessary in very low-density regions. The disc is assumed to be irradiated by the protostar, with a background temperature profile ${T_\mathrm{irr}}^4 =  {(150\,\mathrm{K}(r/\mathrm{au})^{0.5})}^4 + {(3\,\mathrm{K})}^4$, where $r$ is radius from the central protostar. 

\smallskip

The disc is initialized with a flat $Q \approx 1.2$ profile \citep{Toomre1964} and a temperature profile following $T \propto r^{-0.5}$, which gives a surface density profile $\mathit{\Sigma} \propto r^{-1.75}$ for a Keplerian disc. The disc is seeded with a random density perturbation to promote the growth of gravitational instabilities. The initial $Q$ value is low enough to guarantee a very rapid onset of spiral structure, but not so low as to overshoot the instability regime severely. While the I2011 calculations were performed in the context of a massive disc undergoing an outburst of GIs, the simulations here are intended to explore a more protracted phase of instability in a lower mass disc. As such, we evolve the new disc past the outburst phase of unstable discs \citep{Mejia2005} before we begin tracing fluid elements. After about 1290\,yr of evolution (about 5 orbits at 40 au), 2000 fluid elements are placed randomly in the disc, but weighted by mass to reflect the actual distribution of material. This is twice as many fluid elements as used in I2011. The fluid elements are evolved by interpolating the gas flow from the surrounding cells to a given tracer's position and integrated directly. Local disc conditions such as pressure, density, and temperature are also interpolated to the fluid element's position, allowing us to trace the thermodynamic history of the gas. Of the 2000 fluid elements evolved in the simulation, nine are lost through a grid boundary and are not included in the following analysis.

\begin{figure}
    \includegraphics[width = 0.475\textwidth]{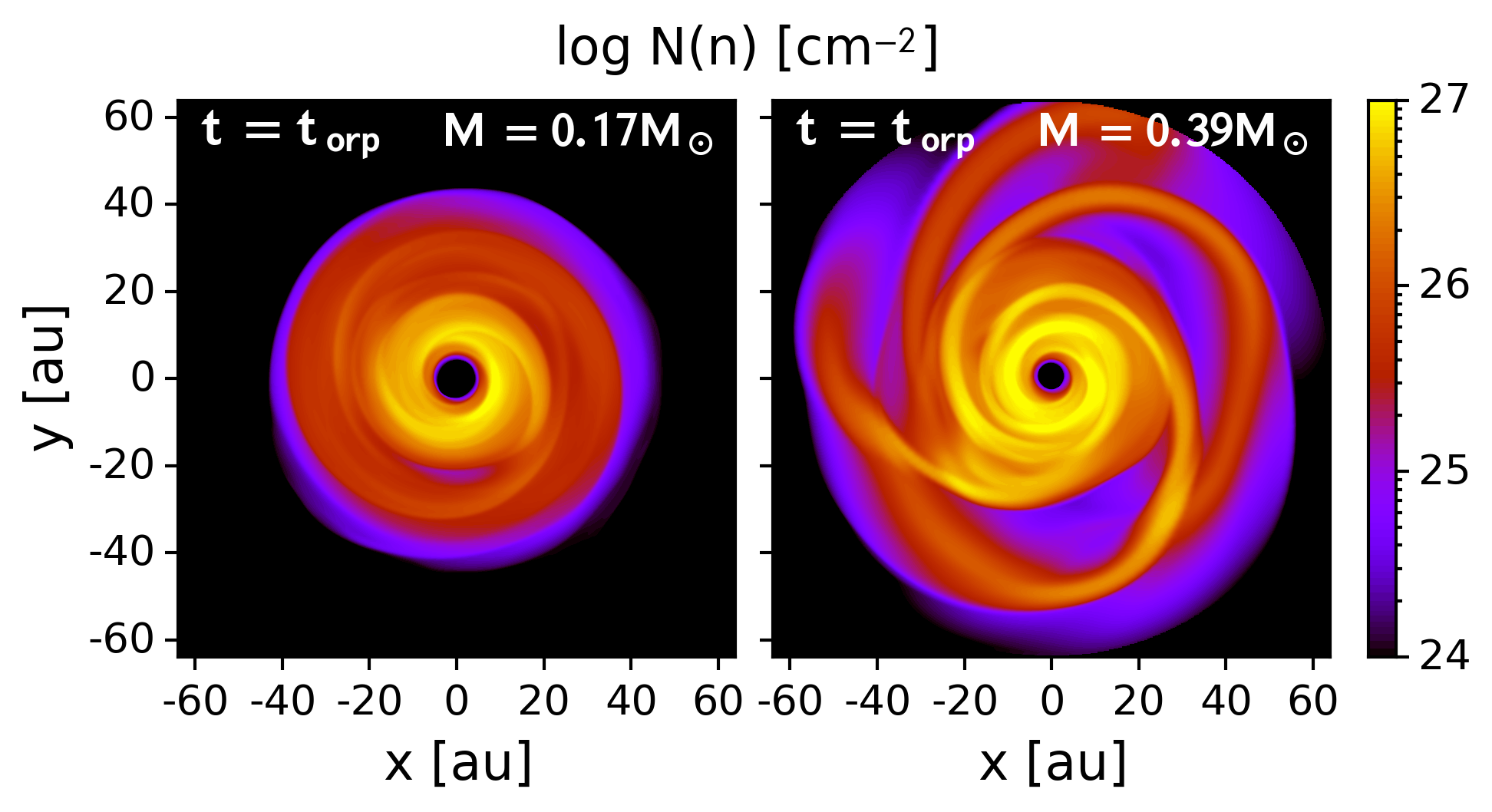}
    \caption{Nuclei column density at $t$ = $t_{\mathrm{orp}}$ for the $0.17\,\mathrm{M}_{\odot}$ (left) and $0.39\,\mathrm{M}_{\odot}$ (right) discs as viewed from above. The more massive disc (right) undergoes a violent instability, while the lower mass disc (left) shows a more protracted phase of instability.}
    \label{fig:NCD440ILEE}
\end{figure}

\begin{figure}
    \includegraphics[width = 0.475\textwidth]{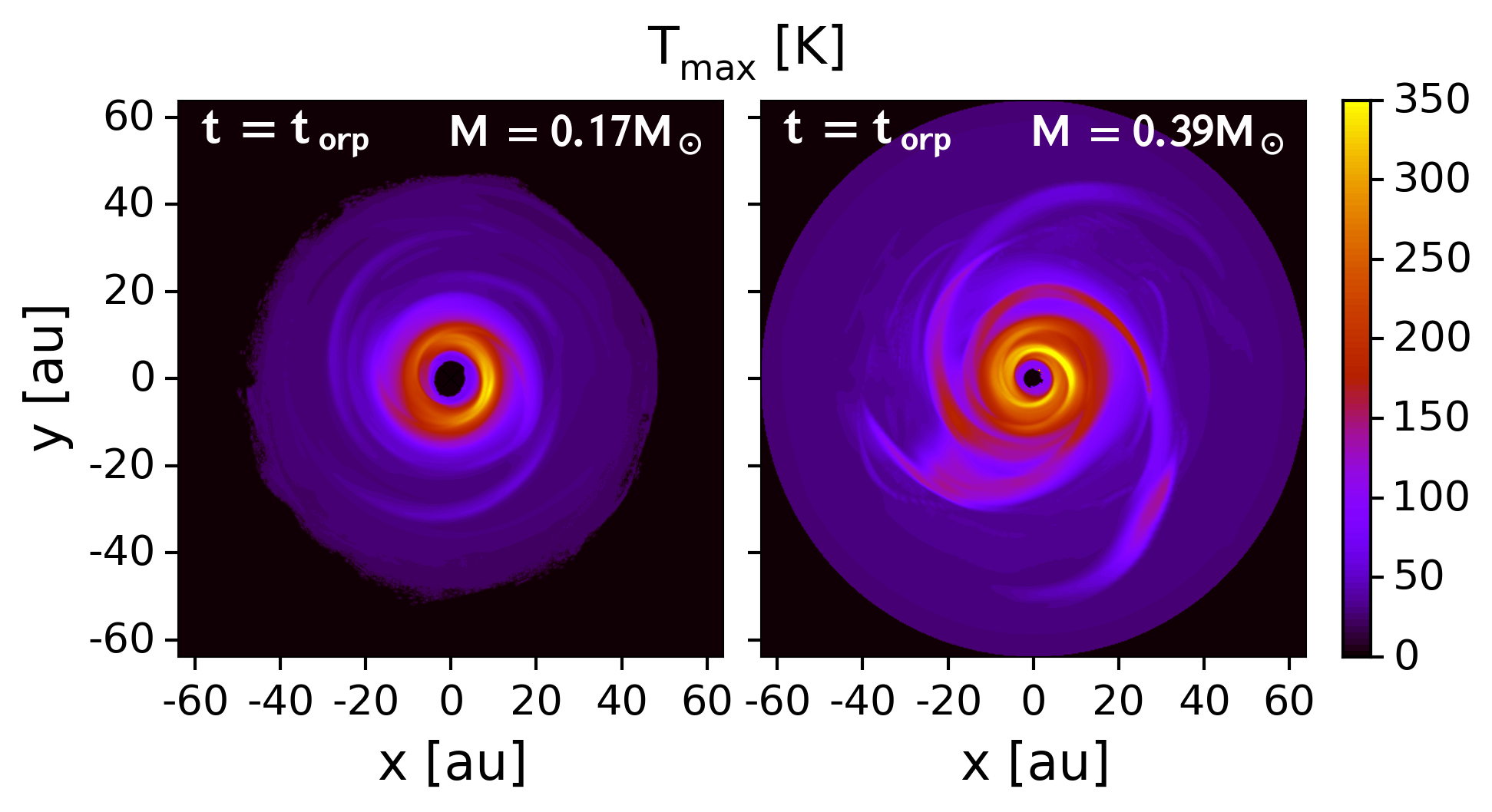}
    \caption{As in Figure \ref{fig:NCD440ILEE}, but showing the maximum line-of-sight temperature.}
    \label{fig:LOST440ILEE}
\end{figure}

\subsection{Comparison with the more massive disc}

The more massive disc featured in I2011 was simulated for 389\,yr, and because of this short duration the hydrodynamical results were repeated for ten cycles, i.e.\ using the final abundances after 389\,yr as initial abundances for the next run, in order to achieve a well mixed chemical distribution. This introduced unrealistic discontinuities into their chemical calculations. Here, we run the new simulation continuously for 2047\,yr with fluid element tracers present, which means we can follow the behaviour of the lower mass disc over a longer period without artificial cycling, allowing for a more self-consistent analysis. 

\smallskip

The more massive disc was initialised with a $Q \approx 1$ profile for most of the disc and a $T_{\mathrm{irr}}$ = $140(r/\mathrm{au})^{0.5}$ + 10$\,\mathrm{K}$ radiation field, requiring a surface density profile $\mathit{\Sigma} \propto r^{-1.75}$. Although these parameter initialisations are similar to the lower mass disc setup, these discs are not directly comparable because they trace different potential phases of protoplanetary disc evolution; the higher mass disc is more applicable to a violent burst in activity, whereas the lower mass disc undergoes a more quiescent protracted period of evolution. However, we can compare the discs after a consistent number of outer rotational periods (ORPs), which defines the time since the inclusion of tracers at a particular radii. At $30\,\mathrm{au}$ the ORP is approximately 140\,yr in the more massive disc and 165\,yr in the less massive disc. We find that the more massive disc has completed $2.7\,\mathrm{ORPs}$ at $30\,\mathrm{au}$ by the end of the simulation, $t = \IleeTime{}\,\mathrm{yr}$. The lower mass disc completes the same number of ORPs at the same radius after approximately $t = \ORPTime{}\,\mathrm{yr}$. Hence, the discs are roughly comparable dynamically at these respective times, which we denote as $t$ = $t_{\mathrm{orp}}$ for both discs, and we use this as the basis for our comparisons.

\smallskip

\begin{figure*}
    \centering
        \includegraphics[width = 1.0\textwidth]{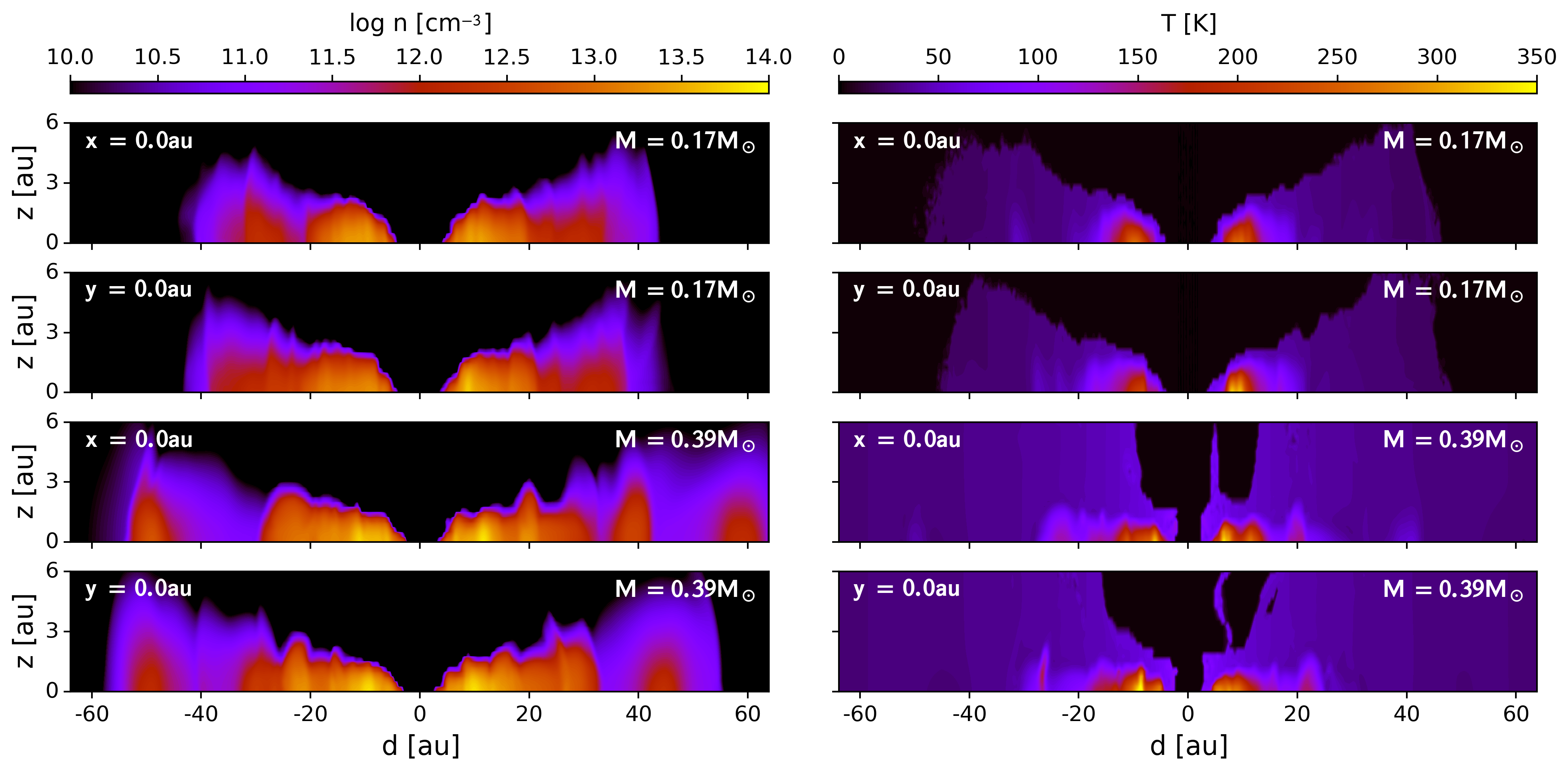}
    \caption{Nuclei number densities and temperatures within slices of the disc interior at $t$ = $t_{\mathrm{orp}}$ for the $0.17\,\mathrm{M}_{\odot}$ (top four panels) and $0.39\,\mathrm{M}_{\odot}$ (bottom four panels) discs. Slices are taken at the specified location and $d$ denotes the distance from the disc centre along the appropriate orthogonal axis.}
    \label{fig:NDTSlices}
\end{figure*}

Figures \ref{fig:NCD440ILEE} and \ref{fig:LOST440ILEE} show the column density of total nuclei and maximum line-of-sight temperature, respectively, within the featured discs at $t$ = $t_{\mathrm{orp}}$, as viewed from above. At this time the maximum temperature and number density within the more massive disc are 400$\,\mathrm{K}$ and $8\times10^{13}\,\mathrm{cm^{-3}}$, respectively, which is 1.2 times hotter and 1.4 times denser than within the less massive disc. Both figures show that prominent spiral features have formed in each disc but the more massive disc is significantly more flocculent due to the stronger gravitational instabilities and evolutionary phase.

\smallskip

Figure \ref{fig:NDTSlices} shows slices of the disc interior along the x = 0\,au and y = 0\,au axes within the less massive and more massive discs at $t$ = $t_{\mathrm{orp}}$. Overall the interior structures are similar; both discs show a flared morphology, with the hottest and densest material located in a ring-like structure approximately 10\,au from the protostar, and possess uneven surfaces resulting from the shocks \citep{Boley2006}. However, the more massive disc is also clearly more extended and features enhanced temperatures at large radii, in contrast to the lower mass disc. Moreover, there is a higher ambient temperature because of the more massive central protostar and disc. These differences in outer disc temperature structure in particular may have a prominent effect on the chemistry as more energy is available for chemical processes and reactions at larger radii.

\smallskip

\begin{figure*}
    \centering
    \includegraphics[width = 0.75\textwidth]{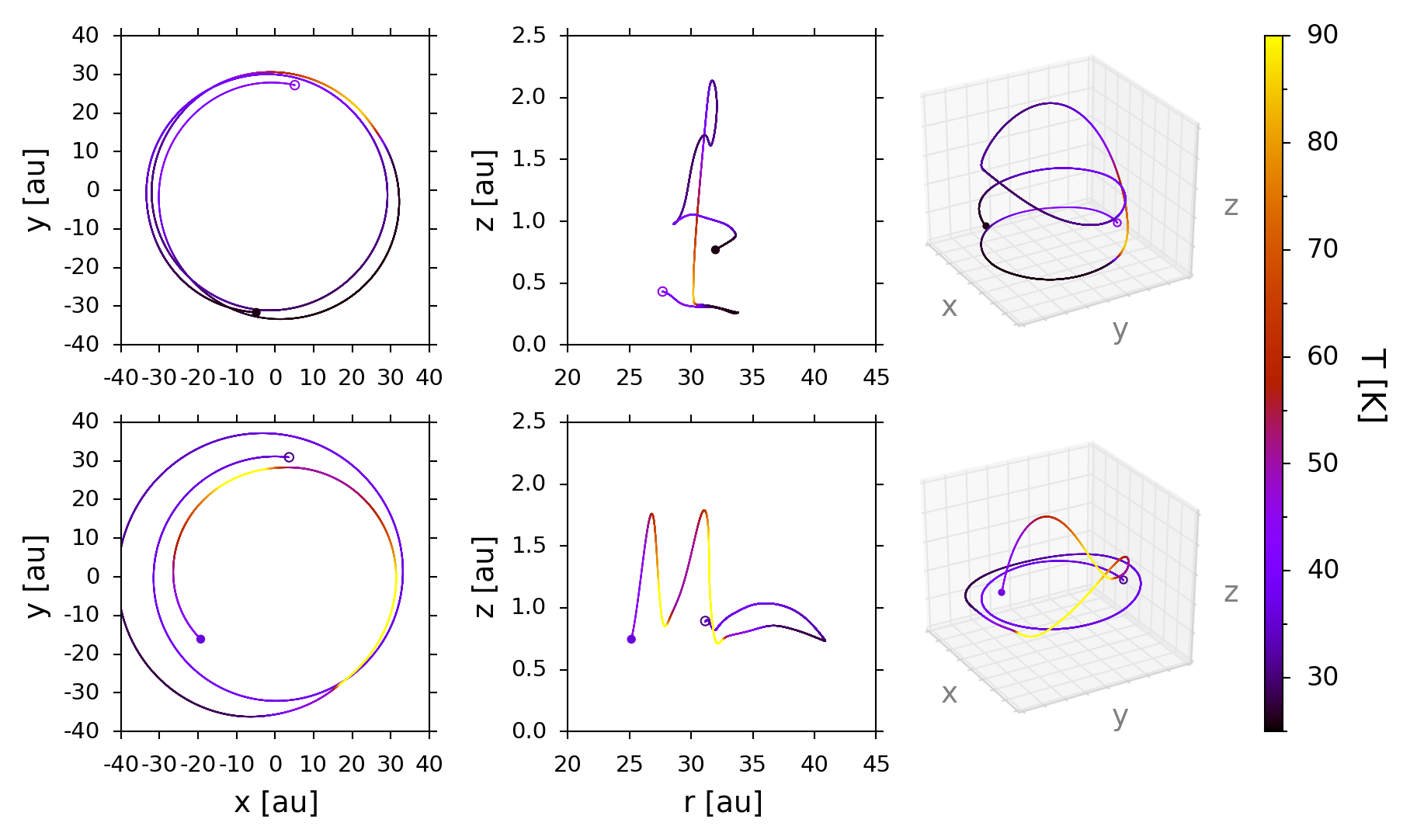}
    \caption{Positions and corresponding temperatures of fluid elements extracted from the $0.17\,\mathrm{M}_{\odot}$ (top) and $0.39\,\mathrm{M}_{\odot}$ (bottom) discs that demonstrate similar orbital behaviour within $2.7\,\mathrm{ORPs}$ at $30\,\mathrm{au}$ in each disc. The open circles denote $t$ = $0$ and the closed circles denote $t$ = $t_{\mathrm{orp}}$.}
    \label{fig:1556225Orbits}
\end{figure*}

\begin{figure}
    \includegraphics[width = 0.475\textwidth]{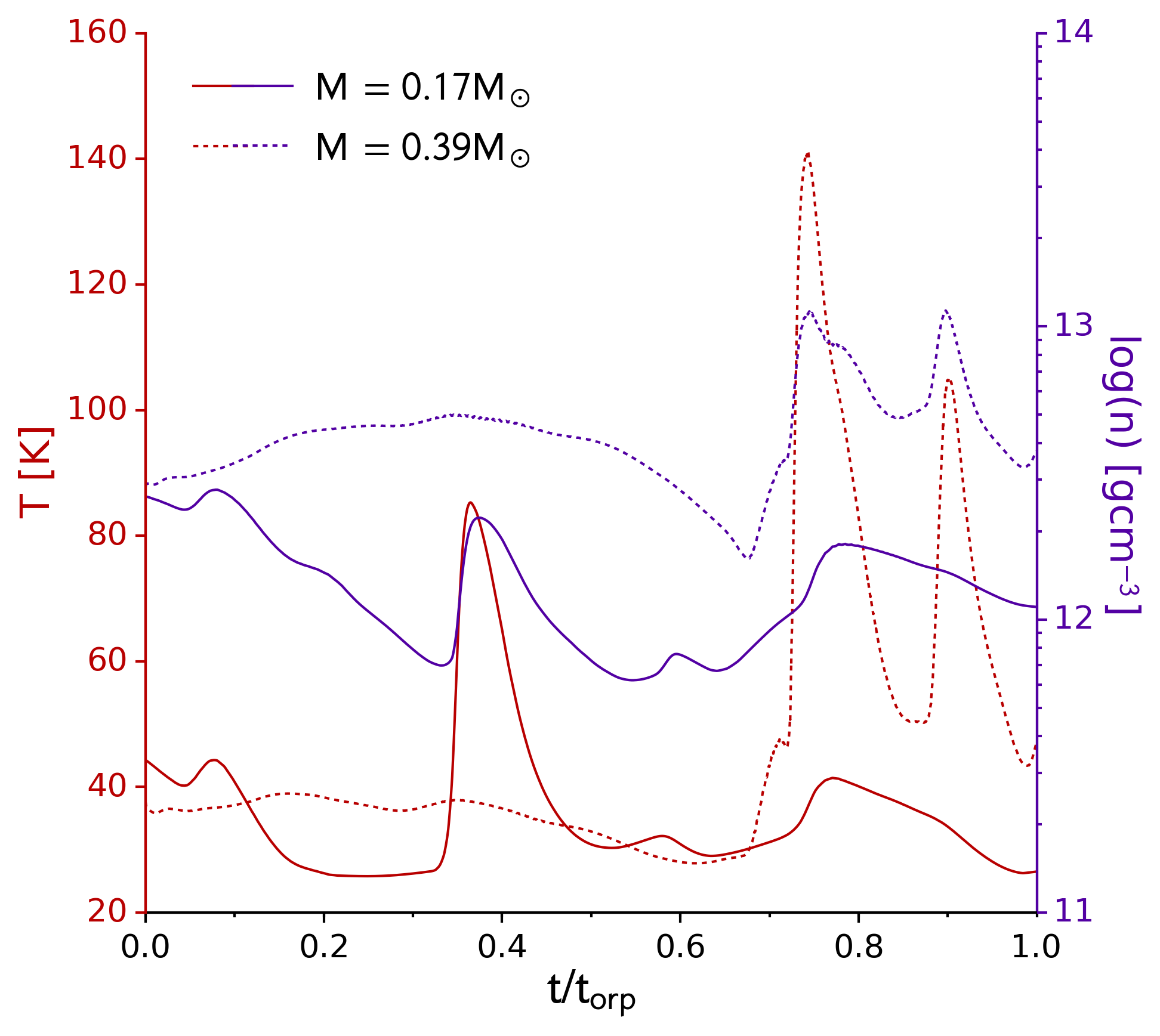}
    \caption{Temperature and number density histories of fluid elements extracted from the $0.17\,\mathrm{M}_{\odot}$ and $0.39\,\mathrm{M}_{\odot}$ discs that follow similar trajectories within $2.7\,\mathrm{ORPs}$ at $30\,\mathrm{au}$ in each disc (see Figure \ref{fig:1556225Orbits}).}
    \label{fig:1556225TempDensity}
\end{figure}

The number densities and temperatures of fluid elements tracked throughout the disc simulation are used as the input for the chemistry model. Figure \ref{fig:1556225Orbits} shows the trajectories of a fluid element in the outer regions of each disc, which is only intended to be illustrative of the types of trajectories that fluid elements can follow. Both fluid elements are initially roughly $30\,\mathrm{au}$ from the centre and orbit on a trajectory that first radially expands and then contracts. We identify the collisional shocks experienced by each fluid element from the concurrent peaks of temperature and number density, seen in Figure \ref{fig:1556225TempDensity}, and pressure (not shown). The shocks cause the fluid elements to be displaced vertically upwards in the discs before falling back towards the midplane. This is an effect seen in gravitationally unstable disc simulations as the strong spiral waves in a vertically stratified disc are both shocks and hydraulic jumps \citep[e.g.][]{Boley2006}.  The velocities of  the shock waves in the disc are relatively low (a usual Mach number  is 2, implying a shock velocity of a few km s$^{-1}$).  Such low Mach numbers are due, in part, to the shallow pitch angles of the spiral arms, which  significantly reduce the speed of a fluid element normal to the spiral shock \citep[see][for a detailed discussion]{Boley2008}.

\subsection{Evolution of the lower mass disc}

\begin{figure}
    \includegraphics[width = 0.475\textwidth]{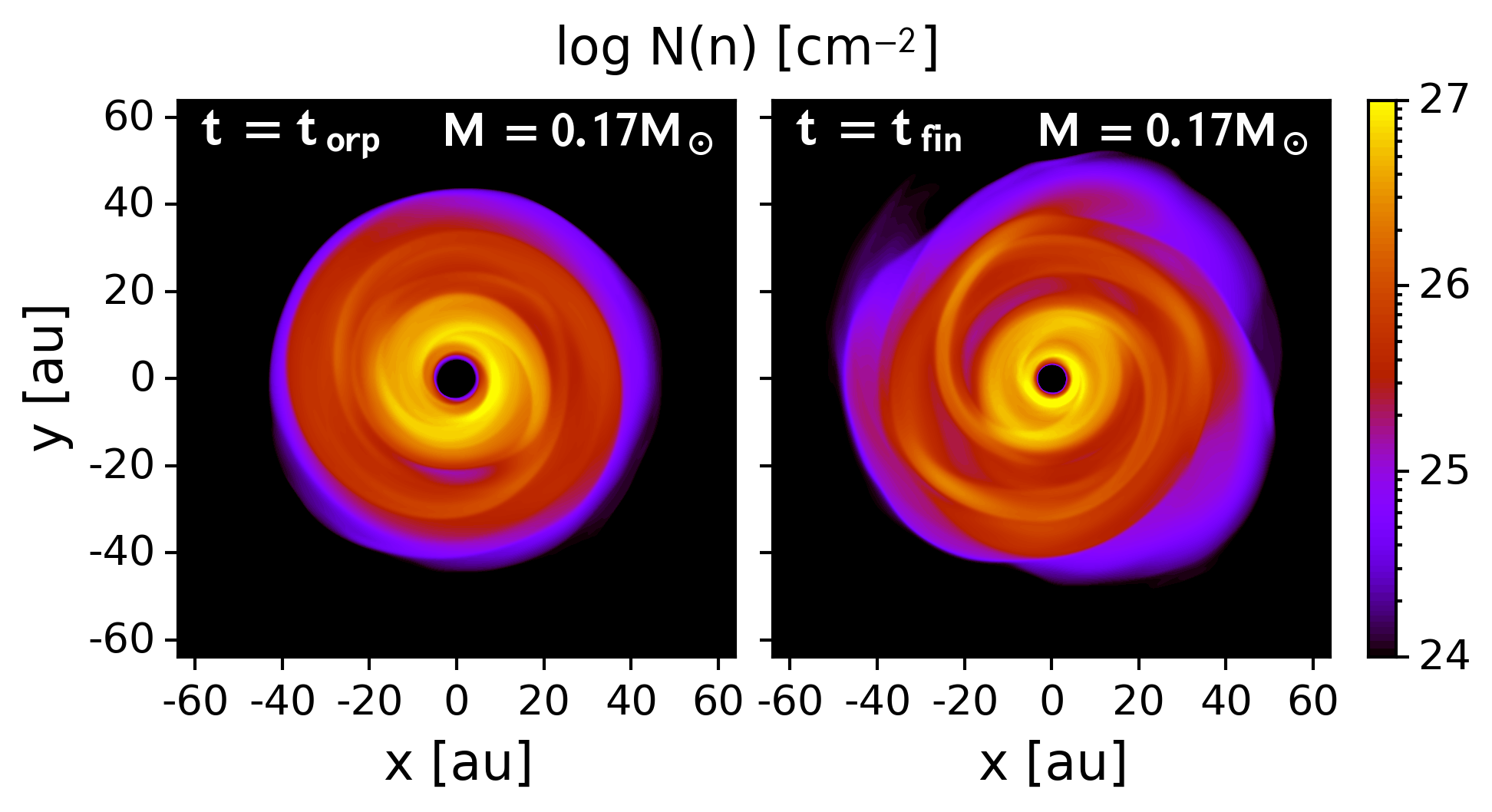}
    \caption{Nuclei column density at $t$ = $t_{\mathrm{orp}}$ (left) and $t$ = $t_{\mathrm{fin}}$ (right) for the $0.17\,\mathrm{M}_{\odot}$ disc as viewed from above. As the disc evolves dynamically, the spiral structure becomes much more flocculent.}
    \label{fig:NCD4402043}
\end{figure}

\begin{figure}
    \includegraphics[width = 0.475\textwidth]{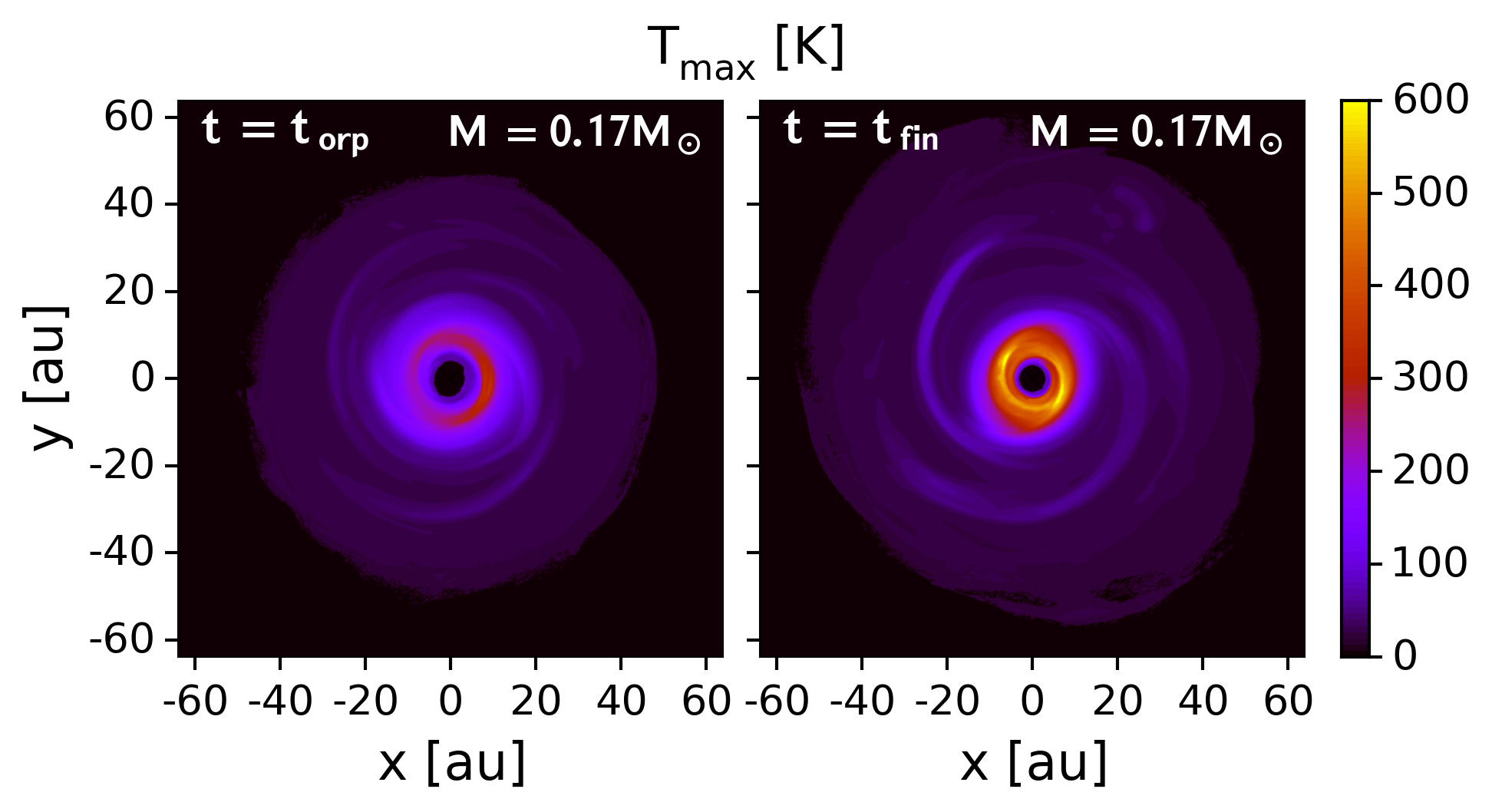}
    \caption{As in Figure \ref{fig:NCD4402043}, but showing maximum line-of-sight temperature. The maximum temperature reached is approximately 600\,K.}
    \label{fig:LOST4402043}
\end{figure}

Figures \ref{fig:NCD4402043} and \ref{fig:LOST4402043} show the nuclei column density and maximum line-of-sight temperature, respectively, within the $0.17\,\mathrm{M}_{\odot}$ disc at $t$ = $t_{\mathrm{orp}}$ ($t$ = $\ORPTime{}\,\mathrm{yr}$) and at the end of the simulation $t$ = $t_{\mathrm{fin}}$ ($t$ = \FinalTime{}\,yr), as viewed from above. Over the duration of the simulation the average mass flux through the disc is $10^{-6}\,\mathrm{M}_{\odot}\mathrm{yr}^{-1}$. By the end of the simulation the disc is clearly more flocculent as the gravitational instabilities have driven progressively denser spiral waves. A video of the lower mass disc evolution demonstrates this and is provided as online supplementary material.

\begin{figure*}
    \includegraphics[width = 0.75\textwidth]{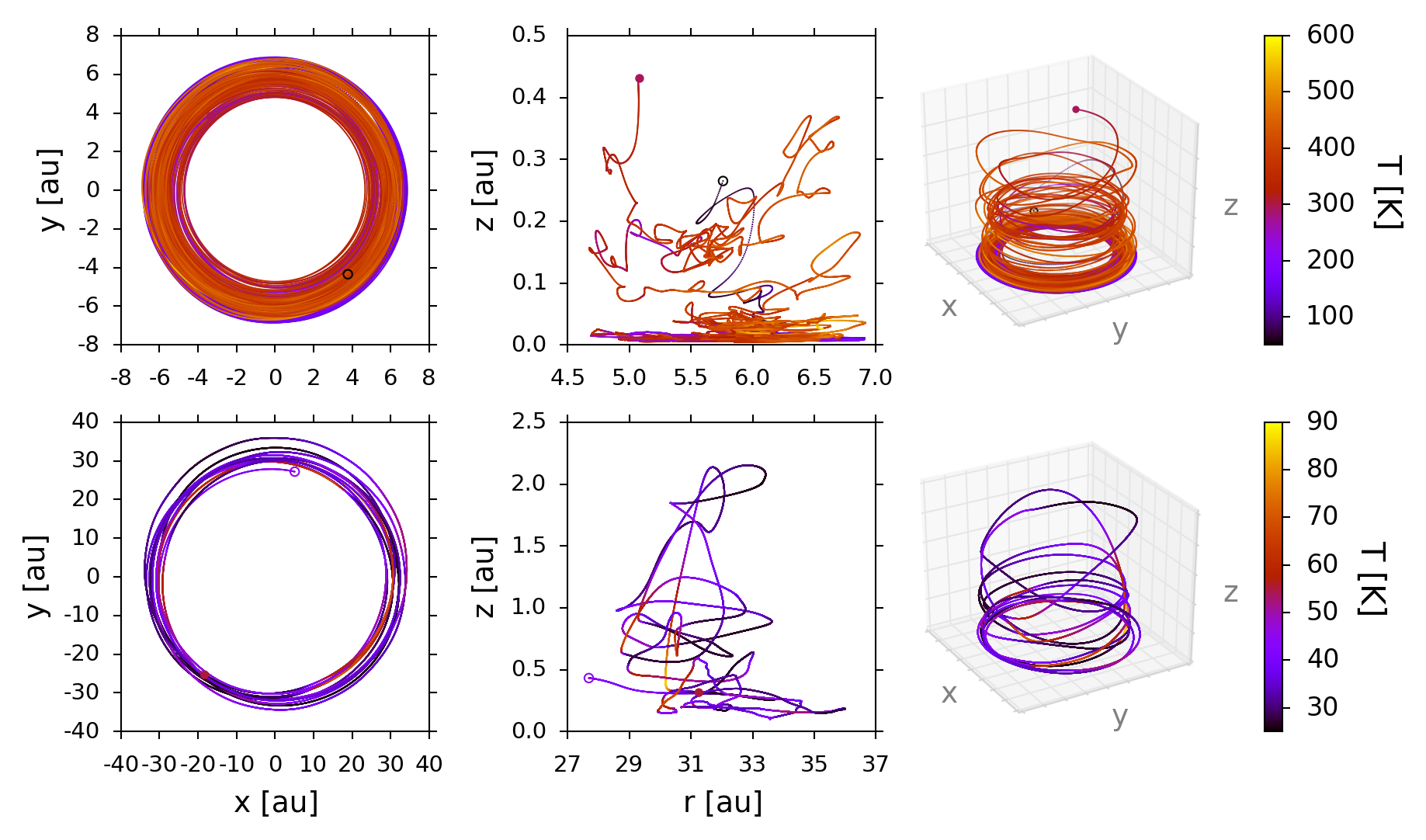}
    \caption{Orbital positions and corresponding temperatures of fluid elements extracted from the inner (top) and outer (bottom) regions of the $0.17\,\mathrm{M}_{\odot}$ disc. The open circles denote $t$ = $0$ and the closed circles denote $t$ = $t_{\mathrm{fin}}$.}
    \label{fig:14331556Orbits}
\end{figure*}

\begin{figure}
    \includegraphics[width = 0.475\textwidth]{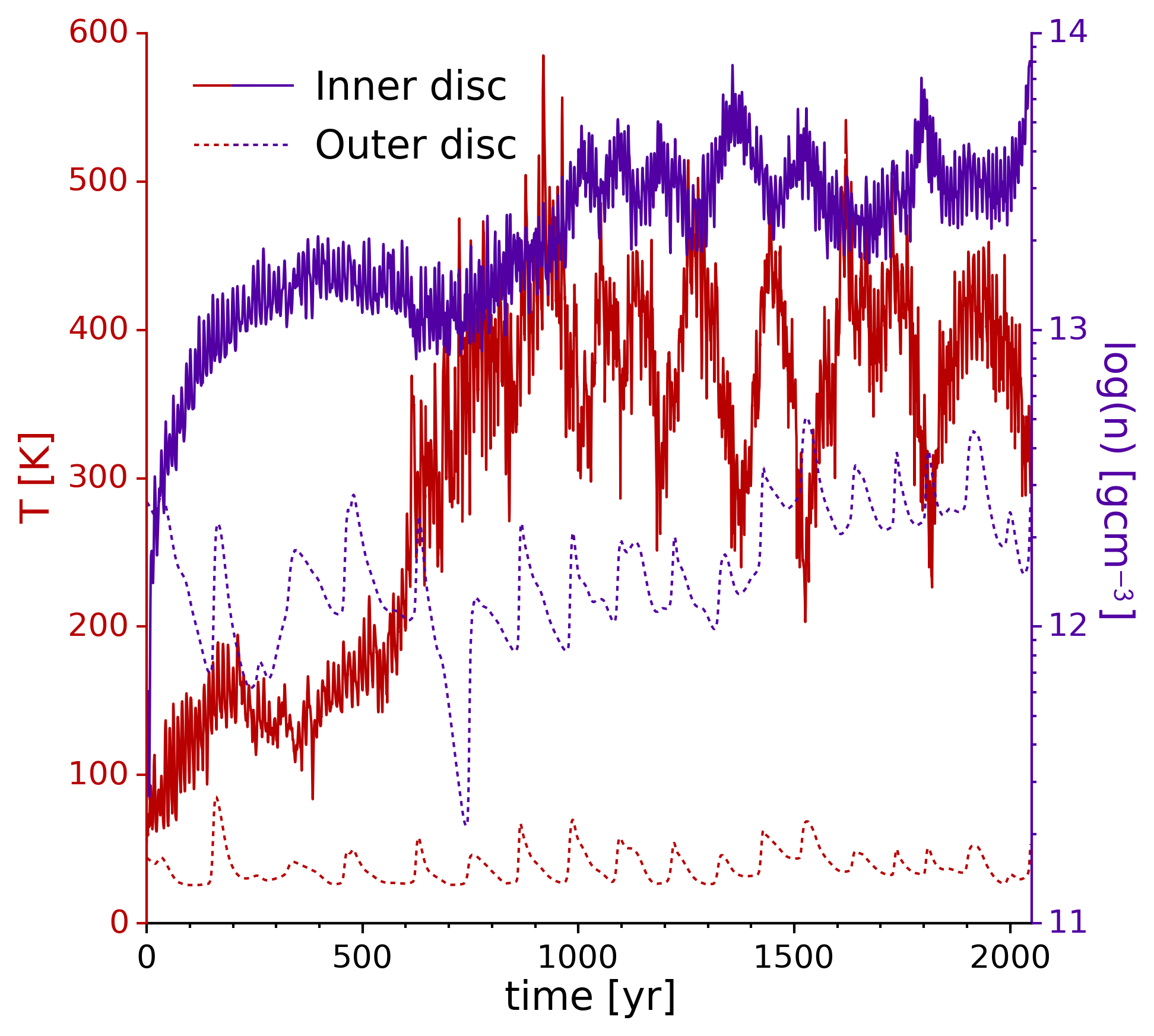}
    \caption{Temperature and number density of fluid elements extracted from the outer and inner regions of the $0.17\,\mathrm{M}_{\odot}$ disc (see Figure \ref{fig:14331556Orbits}). The rapid variations in temperature and number density of are a consequence of collisional shocks, which are much less frequent in the outer disc. The large-scale changes seen only in the inner disc are a result of the fluid element passing through cooler and denser, or more rarefied and hotter regions of the inner disc that are in rough pressure equilibrium.}
    \label{fig:14331556TempDensity}
\end{figure}

Figure \ref{fig:14331556Orbits} shows the trajectories of two fluid elements located within the inner and outer regions of the lower mass disc, throughout the entire duration of the simulation. This behaviour is also shown in the movie provided as online supplementary material (blue trails). The temperature and number density histories of these fluid elements are shown in Figure \ref{fig:14331556TempDensity}, and the outer disc temperature and number density evolution displayed is for the same fluid element as featured in Figures \ref{fig:1556225Orbits} and \ref{fig:1556225TempDensity}. In the outer disc, the temperature profile is determined primarily by the stellar irradiation, with excursions due to spiral shock heating. As a result, over the duration of the entire simulation, the temperature and number density of the outer disc fluid element appear periodic. This period is approximately 100\,yr, which means that the spiral waves propagate faster than the outer disc rotates; the average corotation of the spiral structure appears to be at about 17\,au based on the average mass flux profile. The inner disc fluid element, however, exhibits significantly different behaviour. The much faster rotation of this fluid element, roughly 10\,yr per orbit, results in much more frequent shock encounters, which produces the significantly faster variations in temperature and number density. Moreover, there is a substantial increase in the average temperature and number density for the inner disc fluid element after about $t = 400\,\mathrm{yr}$. This is because the gravitational instabilities drive mass transport and shock heating. Furthermore, within $\approx20\,\mathrm{au}$, the radiative cooling becomes very inefficient as the densest, inner disc regions are more optically thick than the outer disc. In the inner disc fluid element, the rapid variations in temperature and number density, approximately every 10\,yr, reflect passage through spiral structure. The large-scale changes, however, spanning over 100\,yr or more, reflect different regions of the inner disc that are in rough pressure equilibrium, i.e.\ cool and dense regions, or hot and rarefied regions. 

\subsection[Chemical Model]{Chemical Model}
\label{subsec:Chemical Model}

We use the chemical evolution code developed in I2011, which consists of a network of 125 species and 1334 reactions. These reactions are selected from a subset of UMIST95 \citep{Millar1997}, whose rate coefficients have been updated using KIDA\footnote{http://kida.obs.u-bordeaux1.fr/}, and consist of many processes: positive ion-neutral reactions, ionisation by cosmic rays, charge transfer, proton transfer, hydrogen abstraction, radiative and dissociative recombination with electrons, radiative association, neutral exchange, photodissociation and photoionisation, recombination of ions with negative grains, and adsorption and desorption. The code interpolates temperature and number densities extracted from the fluid elements to a higher resolution chemical time-scale using cubic spline fits. The DVODE package \citep{Brown1989} is then used to integrate the rate equations for each considered species, outputting fractional gas-phase abundances, $X_i$ = $n_i$/$n$, where $n$ = $n_H + n_{He} + n_Z$ (see I2011, Equation 1). 

\smallskip

While the chemical model includes various photochemical reactions, we perform the calculations under the assumption that the  disc environment of Class 0 or early Class I objects is well shielded from sources of stellar  and interstellar  radiation, and therefore take $A_{\mathrm{V}}=100\,\mathrm{mag}$.  The combination of envelope infall onto the disc and powerful outflows \citep[see, e.g.,][]{Machida2013} may  drastically limit the illumination of  the upper layers of young discs by the protostellar and interstellar UV field.  However, we allow photochemical reactions induced by cosmic rays, the ionisation rate of which we take to be $\zeta = 10^{-17}$s$^{-1}$.

\smallskip

The chemical model assumes that simple surface chemistry (i.e.  hydrogenation) occurs immediately as species are adsorbed on to the grain surfaces.  In order to simulate any further potential effects of grain surface reactions, our initial species abundances are taken to be representative of observations of cometary ice abundances \citep[see][Table 2]{Ehrenfreund2000} and are given in Table \ref{tab:InitialAbundances}.  Although there is some consistency between cometary and interstellar ices, it is not known whether comets undergo significant chemical processing, and thus, whether they accurately preserve the initial elemental compositions of young systems is not clear \citep[see][Chapter 7]{Caselli2012}.  Therefore, in the future we shall include a detailed treatment of grain surface processes in our chemical model in order to quantify their effect.

\smallskip

As mentioned previously, the velocities of the shock waves in the disc are relatively low (a few km\,s$^{-1}$).  Thus, we do not expect these shocks to cause effects such as the disruption of dust grain cores or sputtering, which typically require shock speeds of at least 25\,km\,s$^{-1}$ \citep[e.g.][]{Caselli1997, VanLoo2013}.  Rather, we focus on the effects of the shocks on the thermal desorption of the ices from the icy mantles of dust grains.

\begin{table}
  \centering
  \caption{Initial fractional abundances $X(i)$ = $n_i$/$n$. Note that a(b) represents a x 10$^b$.}
   \begin{tabular}{cc}
   \hline
    Species  &   Abundance\\
   \hline
    He   &	1.00(-1) \bigstrut[t] \\ 
    H$_{2}$CO  &  1.83(-6)\\
    CO$_{2}$   &  3.67(-5)\\
    CO   &	3.66(-5)\\	 
    HCN   &  4.59(-7)\\
    CH$_{4}$  &	1.10(-6)\\ 
    HNC   &  7.34(-8)\\
    S     &  1.62(-5)\\
    NH$_{3}$  & 	3.30(-6)\\	 
    H$_{2}$S   &  2.75(-6)\\
    SO    &  1.47(-6)\\
    H$_{2}$O  & 	1.83(-4)\\	 
    SO$_{2}$   &  1.84(-7)\\
    OCS   &  3.30(-6) \bigstrut[b] \\
   \hline
   \end{tabular}
   \label{tab:InitialAbundances}
\end{table}

\subsubsection[Modifications]{Modifications}

The chemical evolution code that we use was first presented in the I2011 paper. However, we make some alterations in order to improve its efficiency and success rate. Originally the code used a logarithmic time step in order to capture the fastest initial chemical reactions at a sufficient resolution. This was necessary because the fluid element temperature and density histories in the more massive disc were cycled to compensate for the short time-scale of the physical disc evolution. This introduced artificial numerical discontinuities into the chemical analysis that caused the chemical integrator to fail for a selection of fluid elements. The less massive disc is simulated for a considerably longer time, so we have no need to cycle the chemical abundances. Therefore, we changed the time step to be additive, with a time step of 1000\ seconds so that fast reactions were followed with sufficient resolution throughout the entire disc evolution. The resulting abundances were recorded every 0.05\,yr. Recursion was also incorporated into the chemical code so that a failed integration step was reattempted in a dichotomy paradox fashion, i.e.\ continually halving the time-step, until the integration succeeded or the step size became zero and the run failed. These two adaptations made the code much more robust against rapid changes in temperature and density, which a significant number of inner disc fluid elements possess.

\section{Chemical Results}
\label{sec:Results}

\begin{figure*}
    \includegraphics[width = 1.0\textwidth]{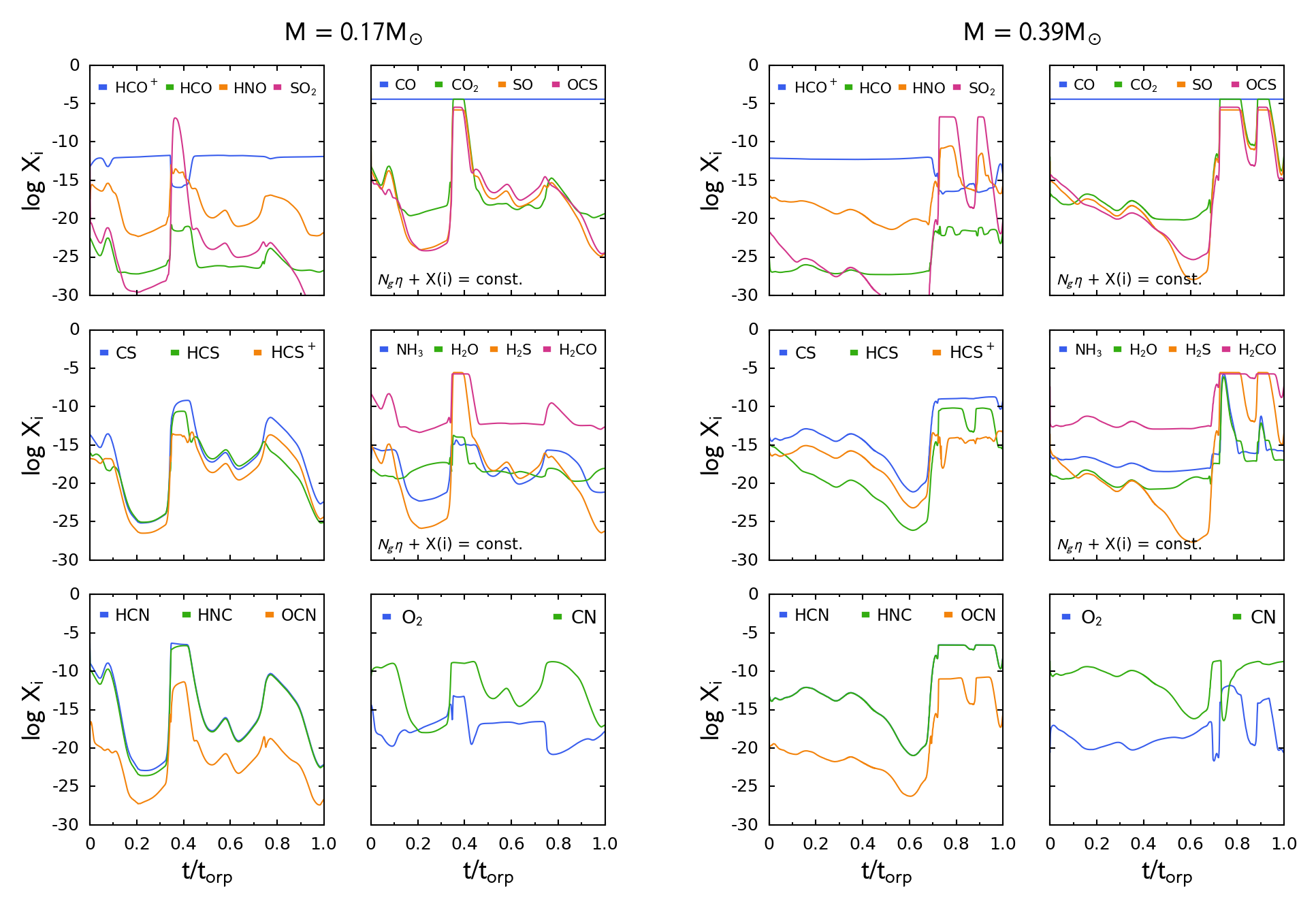}
    \caption{Gas-phase fractional abundances of fluid elements extracted from the $0.17\,\mathrm{M}_{\odot}$ (left) and $0.39\,\mathrm{M}_{\odot}$ (right) discs that follow similar trajectories within $2.7\,\mathrm{ORPs}$ at $30\,\mathrm{au}$ in each disc (see Figure \ref{fig:1556225Orbits}). For each disc, the upper two panels feature species whose total abundance, comprised of the number of molecules on grains and in the gas-phase, is constant over the duration of the simulation.}
    \label{fig:1556225Abundances}
\end{figure*}

\begin{table}
 \centering
  \caption{Maximum fractional gas-phase abundances within the $0.39\,\mathrm{M}_{\odot}$ disc at $t$ = $t_{\mathrm{orp}}$, $0.17\,\mathrm{M}_{\odot}$ disc at $t$ = $t_{\mathrm{orp}}$ and $0.17\,\mathrm{M}_{\odot}$ disc at $t$ = $t_{\mathrm{fin}}$.}
  \begin{tabular}{cccc}
  \hline
   Species  &  \multicolumn{3}{c}{Maximum log $X(i)$} \bigstrut[t] \\
   $i$  &  $0.39\,\mathrm{M}_{\odot}$  &  $0.17\,\mathrm{M}_{\odot}$ & $0.17\,\mathrm{M}_{\odot}$\\
   &            $t$ = $t_{\mathrm{orp}}$  &  $t$ = $t_{\mathrm{orp}}$  &  $t$ = $t_{\mathrm{fin}}$ \bigstrut[b] \\
 \hline
CO          &	-4.4   &  -4.4   &  -4.4 	\bigstrut[t] \\
CO$_{2}$    &	-4.4   &  -4.4   &  -4.4	\\
SO          &	-5.8   &  -5.8   &  -5.8	\\
SO$_{2}$    &	-6.5   &  -6.6   &  -6.3	\\
NH$_{3}$    &	-5.5   &  -5.5   &  -5.5	\\
H$_{2}$O    &	-3.7   &  -3.7   &  -3.7	\\
H$_{2}$CO   &	-5.7   &  -5.7   &  -5.7	\\
H$_{2}$S    &	-5.6   &  -5.6   &  -5.6	\\
HCN         &	-6.4   &  -6.3   &  -6.4	\\
HNC         &	-6.6   &  -6.6   &  -6.6	\\
OCS         &	-5.5   &  -5.5   &  -5.5	\\
HCO$^{+}$   &	-10.8  &  -11.2  &  -10.9	\\
HCO         &	-20.8  &  -20.8  &  -20.3	\\
HNO         &	-7.9   &  -8.1   &  -7.5	\\
CS          &	-8.0   &  -8.1   &  -7.5	\\
HCS         &	-9.6   &  -10.3  &  -9.2	\\
HCS$^{+}$   &	-12.6  &  -12.8  &  -12.1	\\
OCN         &	-10.2  &  -10.3  &  -9.9	\\
O$_{2}$     &	-10.8  &  -11.2  &  -11.2	\\
CN          &	-7.9   &  -7.7   &  -7.2 	\bigstrut[b] \\
  \hline
  \end{tabular}
  \label{tab:MaxAbundancesIleeDynFinal}
\end{table}

\begin{figure*}
    \includegraphics[width = 1.0\textwidth]{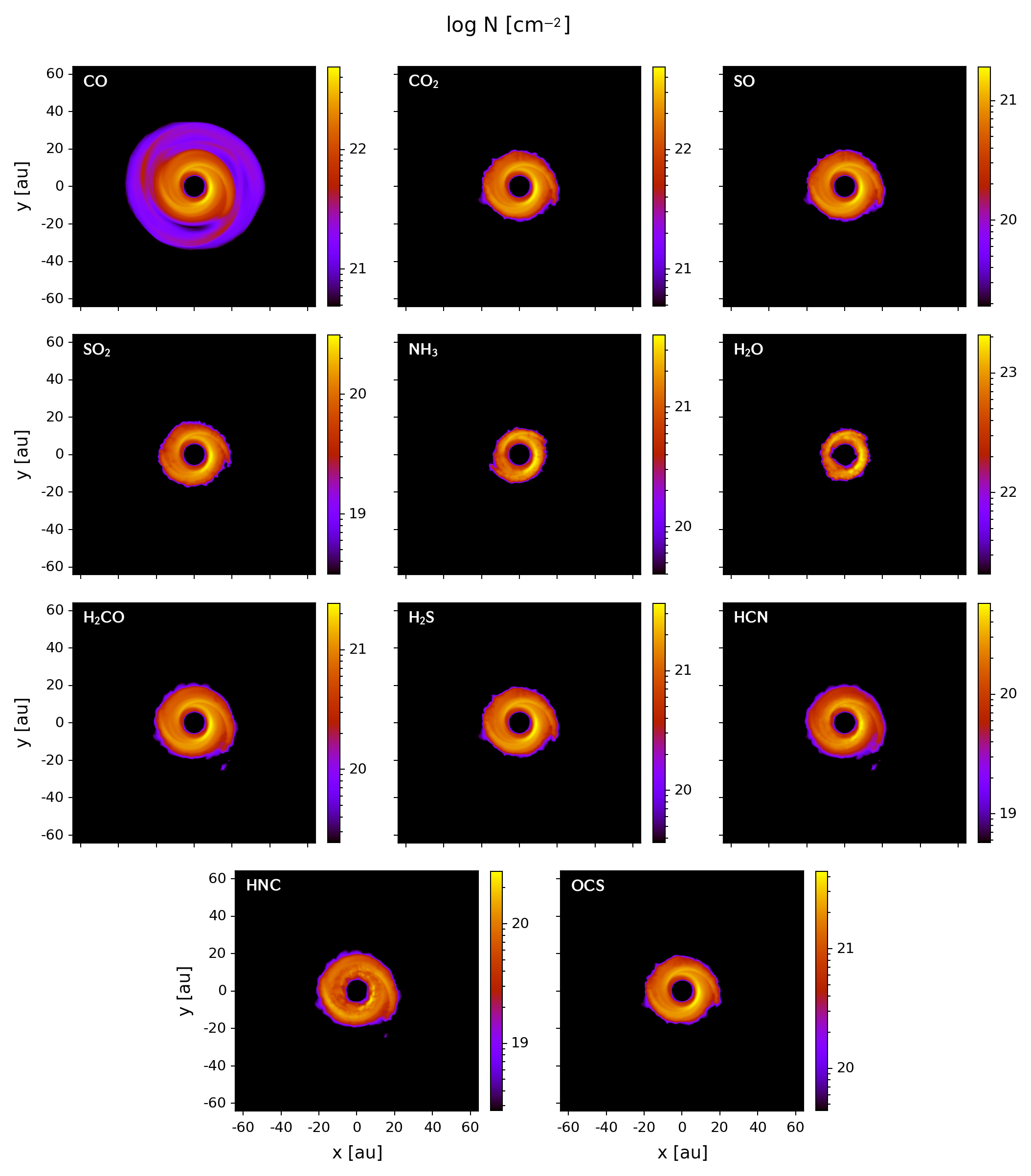}
    \caption{Logarithmic gas-phase column densities of species at $t$ = $t_{\mathrm{orp}}$ with gas-phase fractional abundances determined primarily by thermal adsorption and desorption processes.}
    \label{fig:CD440Constant}
\end{figure*}

\begin{figure*}
    \includegraphics[width = 1.0\textwidth]{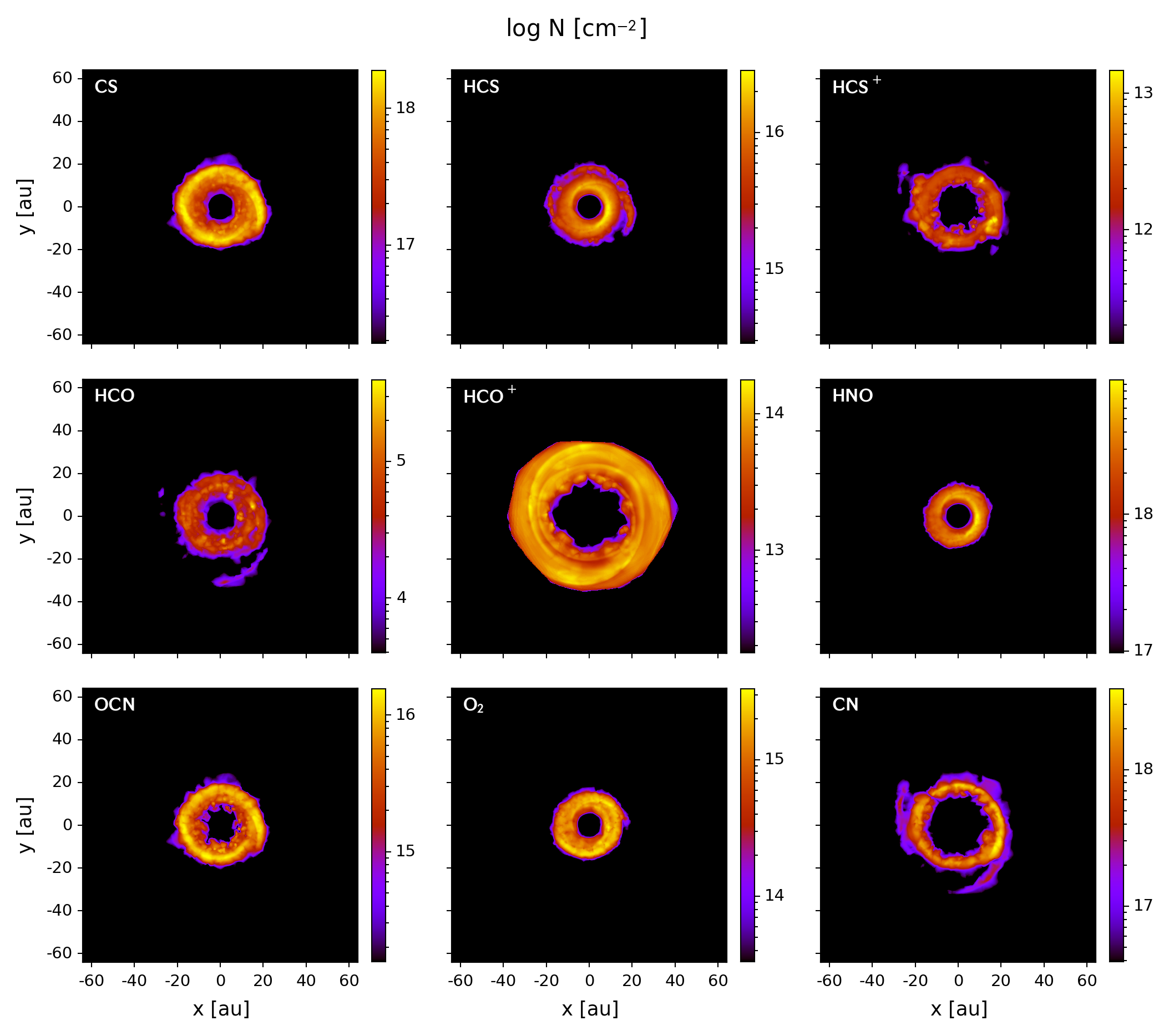}
    \caption{Logarithmic gas-phase column densities of species at $t$ = $t_{\mathrm{orp}}$ with gas-phase fractional abundances significantly affected by chemical reactions.}
    \label{fig:CD440NonConstant}
\end{figure*}

\smallskip

Understanding how gravitational instabilities affect the chemical evolution of the lower mass disc, and how this differs from the more massive disc, is necessary for understanding the potential diversity and observational signatures of such discs. This allows us to determine what tests are the most likely to detect or exclude active instability in any given Class 0 or I source.

\smallskip

Figure \ref{fig:1556225Abundances} shows the fractional gas-phase abundance for 20 species (CO, CO$_2$, SO, SO$_2$, NH$_3$, H$_2$O, H$_2$S, H$_2$CO, OCS, O$_2$, HCO, HCO$^+$, HNO, CS, HCS, HCS$^+$, HCN, HNC, CN and OCN) as a function of $t_{\mathrm{orp}}$ for both discs. The abundances are shown for the same fluid elements featured in Figures \ref{fig:1556225Orbits} and \ref{fig:1556225TempDensity}, which as discussed orbit at approximately $30\,\mathrm{au}$. We have included three species, CO$_2$, CN and H$_2$CO, in addition to those that are featured in the I2011 paper because these are significant species that have recently been detected in protoplanetary discs \citep[e.g.][]{Chapillon2012, Qi2013, Bast2013}. 

\smallskip

The gas-phase fractional abundances of the species in the upper right panels (CO, SO, OCS, NH$_3$, H$_2$O and H$_2$S) are only significantly affected by adsorption and desorption processes because their total abundance, comprised of gas-phase, $X(i)$, and grain abundance, $N_g(i)\eta$, is approximately constant during the disc evolution. Here, $\eta$ is the ratio of the number density of dust grains to the number density of nuclei, and is taken to be $\eta$ = 3.3 $\times$ 10$^{-12}$, while $N_g(i)$ is the average number of molecules of the $i$th species adsorbed on to a grain. However, gas-phase chemical reactions do still occur for these species. For example, the dominant formation reaction at low temperatures for SO is O combining with HS. But at low temperatures ($\approx$ $30\,\mathrm{K}$) SO is almost entirely frozen-out, which means the grain abundance dominates, and given the relatively large initial ice abundance (see Table \ref{tab:InitialAbundances}), the effect of this chemical reaction is indistinguishable in the total SO abundance. Likewise, CO is only frozen-out on to grains below $20\,\mathrm{K}$ and hence it is so abundant in the gas-phase that the effect of the shocks appears negligible. Indeed for species with total fractional abundances above approximately 10$^{-7}$, which is most of the species injected into the disc at initialisation, gas-phase reactions enhanced by shocks are too slow to significantly affect the evolution of the abundances. However, although the total fractional abundance of some species does not depend on shocks, the distribution of gas-phase material is still being affected by spiral arms.

\subsection{Comparison with the more massive disc}
\label{subsec:ChemIlee}

Most of the species reach the same maximum gas-phase abundance for both fluid elements during their respective shocks. This is because, despite the contribution from gas-phase chemical reactions, desorption dominates the formation rates near the desorption temperatures (see Appendix \ref{sec:DesTemps}), which is exceeded for most species during the evolution of both fluid elements. Although the selected fluid elements trace the outer regions of both discs, this result is consistent across the entirety of both discs. Table \ref{tab:MaxAbundancesIleeDynFinal} shows the maximum gas-phase fractional abundances for the 20 species selected at $t$ = $t_{\mathrm{orp}}$, and most species show unremarkable differences between the lower mass and higher mass discs. Therefore, these results suggest that the difference in disc mass and instability strength does not significantly affect the chemistry. Rather, the chemistry is affected merely by the presence of shocks, and perhaps only on a short time-scale. 

\smallskip

In order to understand the features expected to exist in observed gravitationally unstable discs, we generated column density maps for the species discussed at $t$ = $t_{\mathrm{orp}}$. A particular aim of this section is to characterise the affect of system mass on chemical diversity, so these maps are comparable to Figures 7 and 8 featured in I2011.

\smallskip

The column density is defined as
\begin{equation}
   N(i, x, y) = \int{}{}n(i, x, y, z)\,X(i, x, y, z)\,dz.
   \label{eq:ColumnDensity}
\end{equation}
We obtain the number density at each x, y, z using the full hydrodynamics simulation as this affords us the highest resolution possible. We then interpolate the fluid element abundances to the same grid using the {\sc{griddata}} and {\sc{simps}} modules in Python. To obtain results for the full vertical extent of such a disc, we assume that number density and abundances are even functions of disc height, i.e.\ mirrored about the $z = 0$ plane.

\smallskip

Figures \ref{fig:CD440Constant} and \ref{fig:CD440NonConstant} show the column density maps for the selected species at $t$ = $t_{\mathrm{orp}}$, separated by the significance of gas-phase chemical reactions as discussed. Figure \ref{fig:CD440Constant} features the species only affected by adsorption and desorption processes due to their high initial abundances, which is reflected in the column density scales. We have included HCN, HNC and SO$_2$ in this figure because the most dominant reactions, electronic recombination of HCNH$^+$ and SO combining with O or OH, are only prevalent at the lowest temperatures and hence only have a marginal effect on the total abundance. Figure \ref{fig:CD440NonConstant} shows the species whose abundances are significantly affected by chemical reactions during the disc evolution.

\smallskip

The CO column density map is the most extensive, which is expected because the freeze-out temperature is around 20$\,\mathrm{K}$, and this only occurs in the outermost regions of the disc ($>45\,\mathrm{au}$). The relatively abrupt edges seen in this map are primarily due to the limited extent of the fluid elements, as the size of the full simulation is significantly larger than the volume defined by the fluid elements followed throughout the disc. We chose to avoid extrapolating the abundances of species beyond the maximum radial extent of the fluid parcels, as the reliability of the chemical results in this regime can not be assured. 

\smallskip

The H$_2$O map is the most confined and this is because water possesses the highest binding energy to grains of all the species we consider. Hence, water requires the most energy to be efficiently desorbed into the gas-phase, requiring a temperature greater than 150$\,\mathrm{K}$ at these pressures. As can be seen in Figure \ref{fig:LOST440ILEE}, this only occurs for the torus of material approximately 10$\,\mathrm{au}$ from the centre. 

\smallskip

The remaining species with the highest abundances and therefore highest column densities have maps that trace regions of the disc in between these two extremes, dependent on their desorption temperatures. As the highest abundances for most of these species exist in the hottest regions nearest the protostar, the edges of the column density maps effectively define the midplane snow lines in the xy plane. Hence, simulating the molecular emission from the disc chemistry and comparing to observations could provide an indicator to the reliability of our model.

\smallskip

The column density maps in Figure \ref{fig:CD440NonConstant}, for the species that are significantly affected by gas-phase reactions, show some different features compared with the maps shown in Figure \ref{fig:CD440Constant}. Firstly, all of the maximum column densities are much lower. Secondly, CS and OCN appear to have similar extent and structure to species such as CO$_2$ and SO, however, their maximum column density occurs in a torus at approximately 20$\,\mathrm{au}$ from the centre. Furthermore, the HCO$^+$ map spans the entire fluid element distribution but possesses a large inner hole. This is due to HCO$^+$ being primarily formed through a reaction with CO, but destroyed through charge exchange with H$_2$O and reactions with HCN and HNC, all of which are most abundant in the inner disc region. HCS$^+$ is also removed by charge exchange with H$_2$O but reflects the distribution of CS and hence is not as extensive. Aside from HCO$^+$, CN is one of the only species to show clear evidence of spiral structure. While not a component of our initial chemical abundances, CN is initially formed in the gas phase via cosmic-ray induced photodissociation of HCN and HNC \citep[see, e.g.,][]{Gredel1989}. The desorption temperature of CN is around 120$\,\mathrm{K}$, which occurs in these spiral features due to the shock heating. At higher temperatures the reaction $\mathrm{CN} + \mathrm{NH}_3 \rightarrow \mathrm{HCN} + \mathrm{NH}_2$ destroys the molecule, hence producing a very large inner hole. As CN maps areas of moderate temperatures, it could be an ideal tracer for transient heating events in protoplanetary discs, potentially driven by gravitational instabilities.

\smallskip

Although the less massive and more massive discs represent different strengths of GI activity, by using a comparable dynamical age in both systems, defined as $t$ = $t_{\mathrm{orp}}$ (2.7$\,\mathrm{ORPs}$ at 30$\,\mathrm{au}$), we are able to draw some comparisons. Figures 7 and 8 in I2011 show column density maps for the same species we have included in this work, except for CN and CO$_2$. All of the maps included in the I2011 paper show much more defined spiral structure, but this is due to the more massive system driving stronger density waves than in the less massive disc. Despite this there are very few remarkable differences between the sets of column density maps; the proportional extent of species in relation to the total disc size is comparable and the maximum column densities are largely coincident. For example, the CO column density maps show much less spiral structure in the lower mass disc at $t$ = $t_{\mathrm{orp}}$, but this is purely a consequence of the mass and instability strength differences between the discs; the maximum CO abundance is identical in both discs due to the consistent initial conditions and low freeze-out temperature (see Table \ref{tab:MaxAbundancesIleeDynFinal}). There are, however, some distinct differences in the column density maps of HCO$^+$. 

\smallskip

There is a 2.5 times lower peak HCO$^+$ abundance in the lower mass disc, which also manifests itself in the column density maps. This is a consequence of the stronger spiral density waves and larger radius of the more massive disc. HCO$^+$ is formed through a reaction involving CO, which proceeds at the fastest rate in low to moderate temperatures (below approximately 150$\,\mathrm{K}$). The destructive reactions $\mathrm{HCO^+} + \mathrm{HNC} \rightarrow \mathrm{HCN} + \mathrm{H}^+ + \mathrm{CO}$, and more notably, $\mathrm{HCO^+} + \mathrm{HCN} \rightarrow \mathrm{HCN} + \mathrm{H^+} + \mathrm{CO}$ become significant at temperatures around 40-60$\,\mathrm{K}$. In the outer regions of the lower mass disc these reactions are prevalent due to the weak shock strength. However, the shocks are stronger in the higher mass disc, and hence, in the outer regions, these destructive reactions are much less significant. This results in a higher peak HCO$^+$ abundance.

\smallskip

Moreover, the different masses of the discs appear to affect the size of the HCO$^+$ inner hole. At higher temperatures, occurring in the inner disc, H$_2$O destroys HCO$^+$ at a dominant rate, producing an inner hole depleted of the cation. After 2.7$\,\mathrm{ORPs}$, the maximum temperature in the more massive disc is 15 per cent hotter than in the lower mass disc and the extent of the high-temperature region in the inner disc is larger. Therefore, in the lower mass disc, HCO$^+$ is suppressed significantly at radii approximately 1.4 times smaller than in the massive disc (see Appendix \ref{sec:HCOpH2ORadialAbundances}, Figure \ref{fig:H2OHCOpAbundances440ILEE}). This result suggests that HCO$^+$ could potentially be used to characterise disc mass and the strength of instabilities in young systems. 

\begin{figure*}
    \includegraphics[width = 1.0\textwidth]{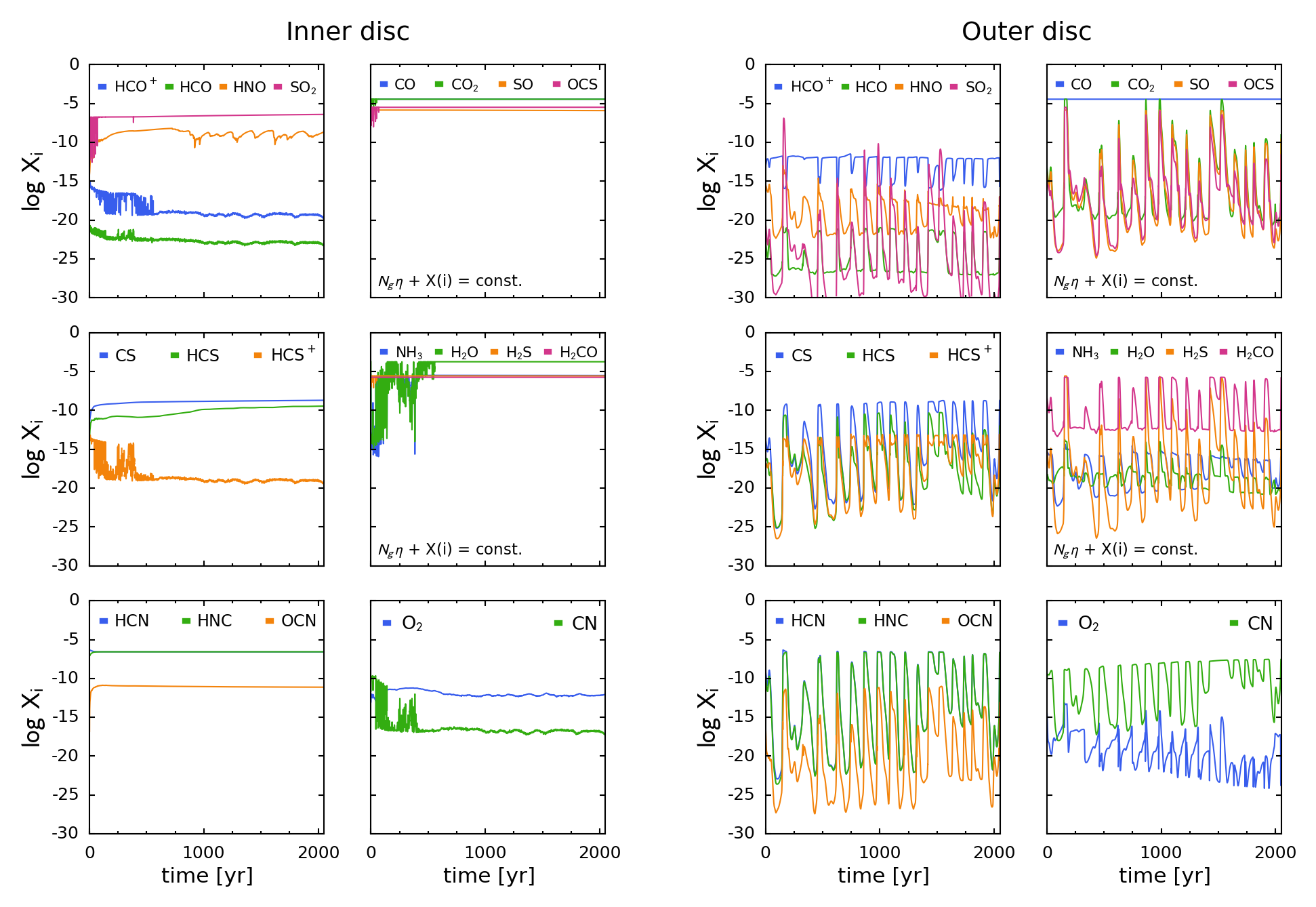}
    \caption{Gas-phase fractional abundances of fluid elements extracted from the inner (left) and outer (right) regions of the $0.17\,\mathrm{M}_{\odot}$ disc (see Figure \ref{fig:14331556Orbits}). For each region, the upper two panels feature species whose total abundance, comprised of the number of molecules on grains and in the gas-phase, is constant over the duration of the simulation.}
    \label{fig:14331556Abundances}
\end{figure*}

The most striking difference, in terms of tracing spiral structure, between our column density maps and those featured in I2011 occurs for HCS. HCS only traces the innermost disc regions in the lower mass disc, whereas, in the more massive disc, HCS also maps the strongest spiral features. More importantly, however, the HCS abundance in these strongly shocked, outer disc regions is equivalent to the peak abundance in the innermost disc; a significant number of other species have distinctly lower abundances in the outer spiral structure. This is because HCS is desorbed above 65$\,\mathrm{K}$ (see Table \ref{tab:DesTempsTable}), but once in the gas-phase, is not destroyed at significant rates. Hence, gas-phase HCS is preserved in shocked regions. As a result, HCS could be a useful tracer of spiral shocks in particularly massive systems.

\subsection{Chemical evolution of the lower mass disc}

By simulating the $0.17\,\mathrm{M}_{\odot}$ protoplanetary disc for multiple dynamical time-scales we are able to investigate how a young, gravitationally unstable system, potentially analogous to the early Solar System, could develop. In addition to the changes in physical properties, such as enhanced maximum density and temperatures as discussed, it is interesting to observe the effects on the chemical evolution of the disc, and in particular characterise whether these are short-term results or can leave a lasting imprint on further chemical evolution.

\begin{figure*}
    \includegraphics[width = 1.0\textwidth]{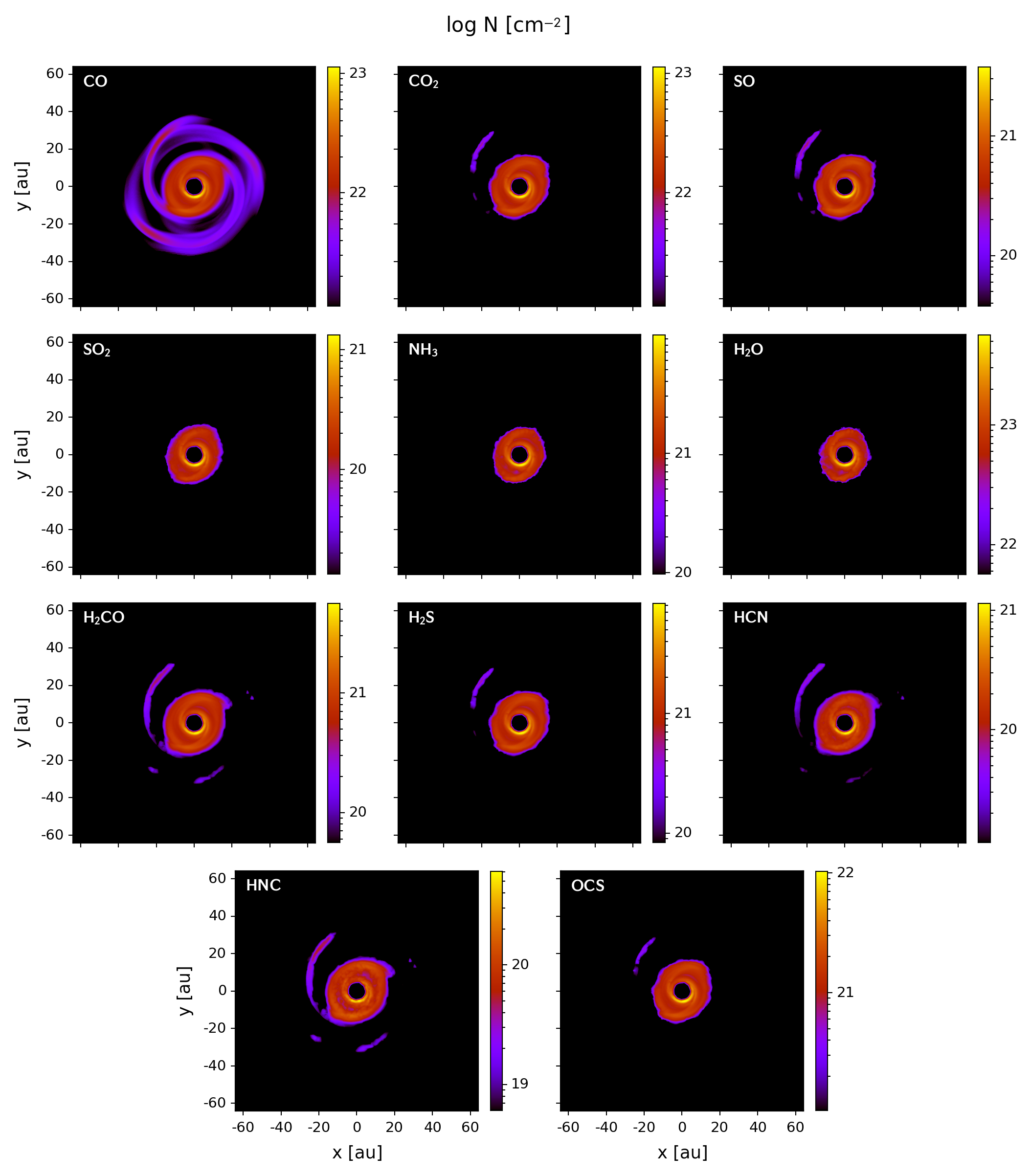}
    \caption{Logarithmic gas-phase column densities of species at $t$ = $t_{\mathrm{fin}}$ with gas-phase fractional abundances determined primarily by thermal adsorption and desorption processes.}
    \label{fig:CD2043Constant}
\end{figure*}

\begin{figure*}
    \includegraphics[width = 1.0\textwidth]{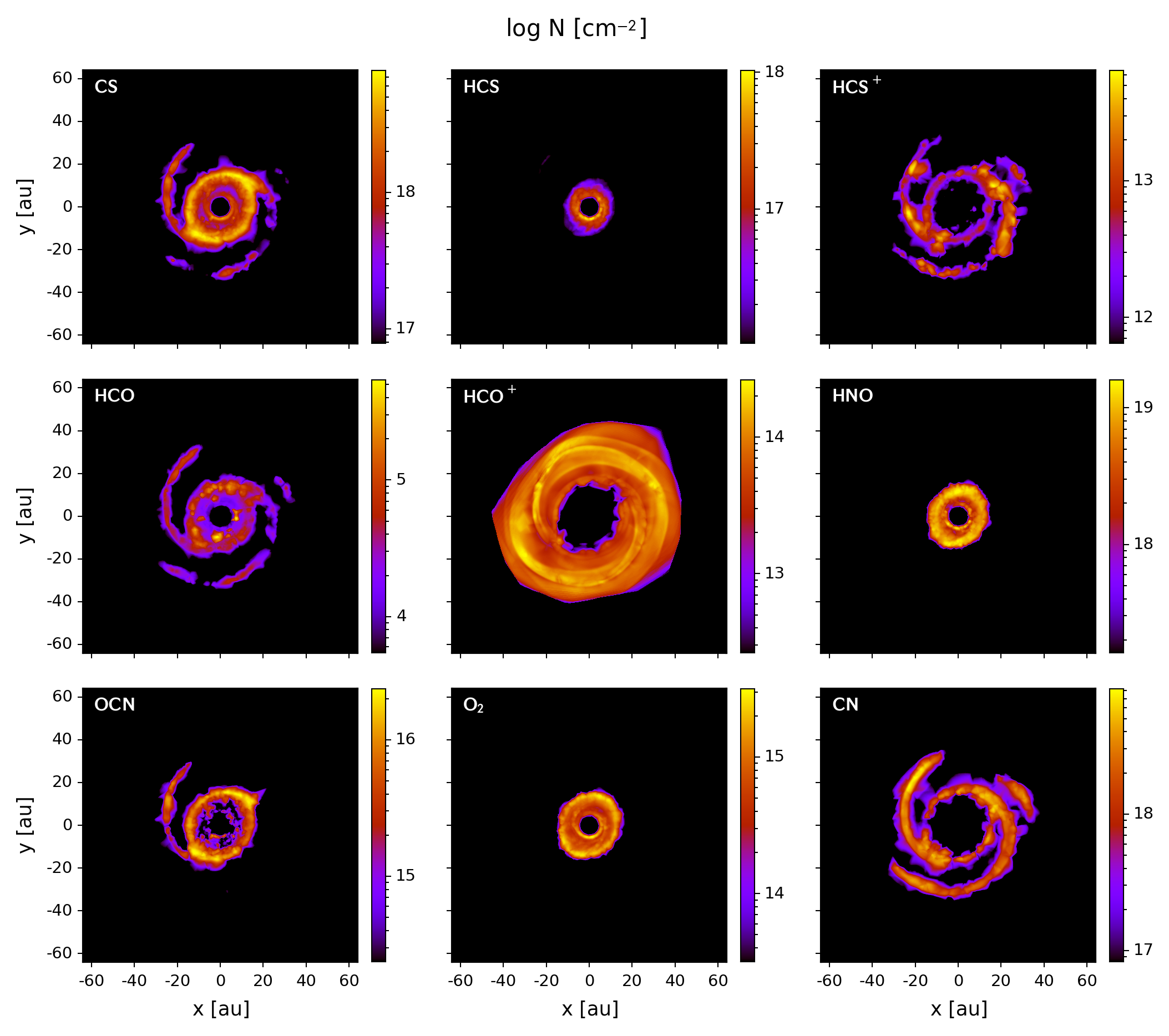}
    \caption{Logarithmic gas-phase column densities of species at $t$ = $t_{\mathrm{fin}}$ with gas-phase fractional abundances significantly affected by chemical reactions.}
    \label{fig:CD2043NonConstant}
\end{figure*}

Figure \ref{fig:14331556Abundances} shows the evolution of the fractional gas-phase abundances of the same 20 species used in Section \ref{subsec:ChemIlee} for fluid elements in the outer and inner regions of the lower mass disc. The fluid elements are the same as featured in Figures \ref{fig:14331556Orbits} and \ref{fig:14331556TempDensity} and orbit at approximately $30\,\mathrm{au}$ and $6\,\mathrm{au}$, respectively. The effects of the shocks on the outer disc fluid element are clearly seen throughout the entire duration of the abundance histories, except for CO due to its very low freeze-out temperature. The shock heating raises the temperature of the fluid element, affording more energy for species to desorb or overcome activation barriers. The exception to this trend is HCO$^+$, which as discussed is destroyed by high abundances of H$_2$O, HCN and HNC. 

\smallskip

The abundance histories of the inner disc fluid element appear significantly different but are entirely consistent with the outer disc abundances. Every species effectively reaches its maximum abundance once the temperature exceeds the energy required for desorption. This occurs at different temperatures, and hence at different times, for each species. For example, HCN and HNC are efficiently desorbed above $60\,\mathrm{K}$, which is instantly reached upon initialisation for the inner disc fluid element. Hence, the fractional gas-phase abundance of HCN and HNC is instantaneously saturated and persists until the end of the simulation. H$_2$O, on the other hand, is desorbed above $150\,\mathrm{K}$, which is only consistently exceeded after around $600\,\mathrm{yr}$. Before this time, the shocks can be seen to have an effect on the abundance as already discussed, and afterwards effectively all the water is in the gas-phase. The behaviour of CN is more complex, but has already been explained. CN is efficiently desorbed above 45$\,\mathrm{K}$ and hence reaches maximum gas-phase abundance nearly instantly. However, at high temperatures CN is formed through cosmic-ray induced photodissociation and destroyed by NH$_3$. Hence, once virtually all of the CN is in the gas-phase, chemical reactions occur at a rate significant enough to affect the abundance. There are more appreciable differences in species for which we have not displayed fractional abundances, but we choose to exclude these as they have not been observed in protoplanetary discs and/or have an insignificantly low abundance. For example, OH has been detected in protoplanetary discs \citep[e.g.][]{Fedele2012, Meeus2012, Fedele2013} but we only recover a peak OH abundance of 10$^{-14}$ because we assume our disc is well shielded from UV photons, which corresponds to a peak column density of 10$^{12}$ cm$^{-2}$ that is likely undetectable. Furthermore, although reactions of atomic  hydrogen can play a destructive role at temperatures greater than $300\,\mathrm{K}$, the  fractional abundance  of atomic hydrogen in  the gas phase  is approximately constant throughout the disc evolution at $\approx10^{-11}$, which corresponds to approximately $1\,\mathrm{ppcm^{-3}}$. As  such, this relatively small  reservoir of  H is quickly  used up and  is not available for further reactions. Other reactions involving O and C atoms play a more dominant role in the chemical evolution.

\smallskip

In order to further assess the implications of GIs on the chemical evolution of young, self-gravitating protoplanetary discs, we produced column density maps at $t$ = $t_{\mathrm{fin}}$ for comparison with Figures \ref{fig:CD440Constant} and \ref{fig:CD440NonConstant}. We used equation \ref{eq:ColumnDensity} and the same procedure and presentation as discussed in Section \ref{subsec:ChemIlee}.
Every species shown in Figures \ref{fig:CD2043Constant} and \ref{fig:CD2043NonConstant} has a larger maximum column density compared to $t$ = $t_{\mathrm{orp}}$, which can be attributed to a combination of enhanced maximum nuclei density and maximum temperature. This suggests, therefore, that despite the fact that some species are destroyed in the inner disc, we still expect to detect higher column densities of all species in a hotter and denser disc.

\smallskip

Investigating the structure of the column density maps at $t$ = $t_{\mathrm{fin}}$ we see that CO is again the most extensive species because of its highly volatile nature. However, due to the more flocculent structure evident in the nuclei column density map at the end of the simulation (see Figure \ref{fig:NCD4402043}), the spiral structure is much more prominent. This strengthening of spiral features is also observed in species with relatively low desorption temperatures, such as CO$_2$, H$_2$CO and CS, but species with the highest binding energy to grains, such as H$_2$O and NH$_3$ still only trace the innermost regions of the disc. CS in particular appears to be a useful molecule in tracing the spiral features expected in young, self-gravitating discs but only after the spiral waves have fully developed. CN, on the other hand, could prove to be excellent in characterising instabilities in embedded discs as it traces non-axisymmetric features throughout the entire disc evolution with a relatively high column density, and, more importantly for current observational constraints, in the outer regions of the disc.

\subsubsection{Persistent effects of shock heating} 

The shock heating in the disc clearly has an instantaneous effect on the chemistry, but it is worthwhile to investigate whether these effects persist. In the inner disc, once the desorption temperature is exceeded for a particular species, the abundance appears unaffected by further shocks. However, Table \ref{tab:MaxAbundancesIleeDynFinal} shows that some species have enhanced maximum abundances at the end of the simulation, e.g.\ about a 10 times increase in maximum abundance of HCS, despite the high temperatures and densities of the inner disc at $t$ = $t_{\mathrm{orp}}$. This suggests that the enhanced densities and temperatures of the inner disc regions driven by GIs in a young, massive disc can lead to subtle changes in the chemical composition. The inner regions of discs (i.e.\ within $20\,\mathrm{au}$), however, particularly in young, embedded systems, are very challenging to observe.

\smallskip

In the more observationally accessible outer disc, the shocks have a continual effect on the chemical abundances due to the low ambient temperature. Although most species appear to reach the same abundances post-shock as pre-shock, we investigated further and found evidence to suggest that consistent shock-heating changes the chemical composition, at least over the short timescale of our simulation. To do this we took a fluid element in the outer disc and fitted a low and high-order polynomial to binned minima of the temperature and number density (see Figure \ref{fig:ShockLHOSTempDensity}). This allowed us to investigate the chemical evolution of an artificial fluid element more appropriate for a trajectory within a disc absent of shocks in the outer regions. 

\begin{table}
 \centering
  \caption{Fractional gas-phase abundances at $t$ = $2000\,\mathrm{yr}$ for a fluid element in the outer disc with shocks (WS), and for the same fluid element without shocks via fitting a low-order (WOSLO) and high-order (WOSHO) polynomial to the temperature and number density minima. Note that we present the results for $t$ = $2000\,\mathrm{yr}$ because at the end of the simulation the fluid element is encountering a shock.}
  \begin{tabular}{cccc}
  \hline
   Species  &  \multicolumn{3}{c}{log $X(i)$} \bigstrut[t] \\
   $i$  &  WS  &  WOSLO  &  WOSHO  \bigstrut[b] \\
 \hline
    CO  &  -4.4  &  -4.4  &  -4.4  \bigstrut[t]\\
    CO2  &  -19.1  &  -17.3  &  -17.4\\
    SO  &  -21.7  &  -19.7  &  -19.7\\
    SO2  &  -27.3  &  -24.5  &  -24.4\\
    NH3  &  -20.3  &  -24.8  &  -24.8\\
    H2O  &  -17.5  &  -16.2  &  -16.2\\
    H2CO  &  -12.9  &  -11.9  &  -11.9\\
    H2S  &  -23.1  &  -21.3  &  -21.4\\
    HCN  &  -23.8  &  -24.1  &  -24.1\\
    HNC  &  -23.7  &  -24.8  &  -24.9\\
    OCS  &  -21.3  &  -20.1  &  -20.1\\
    HCO+  &  -11.8  &  -11.5  &  -11.5\\
    HCO  &  -26.5  &  -25.0  &  -25.1\\
    HNO  &  -21.2  &  -24.3  &  -24.2\\
    CS  &  -22.5  &  -20.9  &  -21.0\\
    HCS  &  -22.1  &  -20.8  &  -20.9\\
    HCS+  &  -23.6  &  -22.2  &  -22.3\\
    OCN  &  -27.8  &  -28.3  &  -28.3\\
    O2  &  -18.0  &  -15.2  &  -15.2\\
    CN  &  -18.0  &  -21.3  &  -21.3  \bigstrut[b]\\
  \hline
  \end{tabular}
  \label{tab:ShockLHOSAbundances}
\end{table}

We found that the abundances of most species are not affected significantly when shocks are removed, as the abundances near the end of the simulation are comparable, as shown in Table \ref{tab:ShockLHOSAbundances}. Hence, the chemical changes for most species are not persistent. However, HNO, CN and NH$_3$ have significantly higher abundances near the end of the simulation when the fluid element is consistently shock-heated in the outer disc. For example, devoid of shocks, HNO is formed via $\mathrm{O} + \mathrm{NH}_2 \rightarrow \mathrm{HNO} + \mathrm{H}$ at an approximate rate $10^{-31}\,\mathrm{cm^{-3}s^{-1}}$. In shock-heated regions, however, this rate reaches a maximum of $10^{-22}\,\mathrm{cm^{-3}s^{-1}}$, and the frequency of shocks results in a limited minimum rate of $10^{-28}\,\mathrm{cm^{-3}s^{-1}}$. This leads to a HNO abundance enhanced by a factor of $10^{3}$. NH$_3$ is enhanced by a factor of $3\times10^{4}$ due to this same phenomenon, only the rate of the reaction $\mathrm{NH}{_4}^+ + \mathrm{e}^- \rightarrow \mathrm{NH}_3 + \mathrm{H}$ is consistently enhanced at least $10^{4}$ times from shock-heating. CN is enhanced by a factor of $2\times10^{3}$ in much the same manner, except that desorption is the dominant formation reaction due to the lower desorption temperature of CN (see Table \ref{tab:DesTempsTable}). 

\smallskip

Shocks in a relatively low-mass, protosolar-like disc are seen to cause persistent changes in the abundance of some species. This has several implications for future studies of young protoplanetary discs. Firstly, whether this effect is independent of the initial conditions should be investigated, because subtle differences in early-phase composition may not be preserved through a short evolutionary period, resulting in similar abundances at the end of the GI phase across a variety of young systems. Secondly, if gravitational instabilities are a significant phase of early disc evolution, simulations of more evolved systems should consider the processing of GIs before setting their initial abundances, in order to account for the enhancement in some important species. Ideally, models of protoplanetary disc chemistry should use a self-consistent approach concerning the molecular core collapse phase \citep[e.g.][]{Visser2011}, followed by GI-driven evolution if the disc-to-star mass ratio is large enough. The resulting abundances can then be used as initial conditions for the simulation of a more evolved, low-viscosity or passive disc, at which point an axisymmetric $\alpha$-disc prescription becomes appropriate. Furthermore, these results suggest that observations of discs could still be used to characterise disc dynamics even if they can not be resolved spatially, which is an important finding. This is because successive shocks enhance the abundances of some important species, and so an observation of a species such as CN, that may otherwise be undetectable in the outer regions of a more passive disc, indicates an enhancement in temperature that could be produced by GIs. Moreover, even if the effects of shock heating are short-lived for some species, the enhancement in abundance is very significant, e.g.\ up to a factor of 10$^{20}$ for SO$_2$ (see Figure \ref{fig:14331556Abundances}), Therefore, detections of transient species may still be indicative of GIs. Further investigation is warranted in this area to determine how permanently gravitational instabilities affect disc composition over longer timescales. 

\subsection{Comparison with axisymmetric models}
\label{subsec:Comparisons}

Comparing the results reported in this paper to other results in the literature is not straightforward due to the varying simulation dynamics, techniques and assumptions. We have used the chemical network developed in I2011, yet even a comparison with this work is complicated because the less massive and more massive discs represent different stages of GI activity and occupy different parameter spaces. Nevertheless, Table \ref{tab:MaxAbundancesComp} shows maximum abundances taken from two recent studies, \citet{Walsh2010} (hereafter W2010) and \citet{Akimkin2013} (hereafter A2013), to provide a very rough comparison to the maximum abundances at the end of our lower mass disc simulation.

\begin{table}
 \centering
  \caption{Maximum gas-phase abundances reported within our work, W2010 and A2013. A dash represents a missing reported value for that species in the corresponding work.}
  \begin{tabular}{cccc}
  \hline
   Species  &   \multicolumn{3}{c}{Maximum log $X(i)$} \bigstrut[t] \\
   $i$  &  This work & W2010 & A2013 \bigstrut[b] \\
 \hline
CO   &	-4.4  &	-4  & -4.0 \bigstrut[t] \\
CO2  & 	-4.4  &	-6  & -\\
NH3  & 	-5.5  &	-   & -4.7\\
H2O  & 	-3.7  &	-4  & -5.0\\
H2CO & 	-5.7  &	-9  & -9.0\\
HCN  & 	-6.4  &	-7  & -8.5\\
HCO+ &	-10.9 &	-6  & -\\
CS   &	-7.5  &	-8  & -\\
O2   &	-11.2 &	-   & -4.5\\
CN   &	-7.2  &	-7  & -\\
C2H  &  -9.9  & -7  & -\\
N2H+ &  -18.9 & -11 & -11.0\\
OH   &  -13.9 & -4  & - \bigstrut[b] \\
  \hline
  \end{tabular}
  \label{tab:MaxAbundancesComp}
\end{table}

We extracted the peak abundances from W2010 by visually inspecting their `fiducial' model between $10\,\mathrm{au}$ and $50\,\mathrm{au}$, which introduces some uncertainty into the quoted values. Nevertheless, there are significant differences between our results that arise from a multitude of factors. Firstly, W2010 model a 1$\,\mathrm{Myr}$ old laminar model disc, which is more evolved than the system we consider and features only local mass transport. As a result their disc possesses the vertically inverted temperature structure observed in Class II YSOs due to the photon-dominated region near the disc surface. Therefore the highest abundances for most species in W2010 occur in a `transition' layer confined to z/r $\approx$ 0.3, where UV and X-ray dissociation is pertinent, which explains their greatly enhanced OH abundance. If we compare only within the midplane regions we recover a similar peak OH abundance. This is also true for the HCO$^+$ abundance, which is expected to be due to the ubiquity of gas-phase CO in both models. The midplane abundance of H$_2$CO in W2010 is much lower than we report due to the more efficient freeze-out in their colder midplane and lower initial abundance; W2010 use significantly different initial abundances, focusing on oxygen-rich low-metallicity elements rather than cometary ices. Due to the strong dependence on initial abundances for some species, those that are only significantly affected by adsorption and desorption processes, we find a higher abundance at the end of our simulation. Furthermore, we do not include atomic N initially in our chemical network which results in a significantly lower N$_2$H$^+$ than W2010 found.

\smallskip

The maximum abundance values quoted for A2013 were determined from figures showing the abundances at $10\,\mathrm{au}$ and $50\,\mathrm{au}$ within their disc model. We use different initial abundances in our model than A2013, who use oxygen-rich low-metallicity abundances similar to W2010, which helps explain our enhanced values for H$_2$O, H$_2$CO and HCN. Contrary to A2013, we do not include N or N$_2$ in our initial conditions and furthermore, A2013 incorporate grain growth in their simulation, which they report enhances the abundance of nitrogen atoms in the midplane. Hence, we record a lower N$_2$H$^+$ abundance. Interestingly, however, A2013 report the same peak N$_2$H$^+$ abundance as W2010 despite the lack of grain growth treatment in the latter. Our lower maximum abundance of O$_2$ is a consequence of our model containing 1.3 times less elemental oxygen and the A2013 model including UV and X-rays that enhance the abundance of OH in the upper disc, which subsequently leads to synthesis of molecular oxygen via a neutral-neutral reaction with atomic oxygen. As the A2013 disc follows the classic $\alpha$ prescription with an array of photoprocesses incorporated into their chemical model, they recover a heated layer in the disc possessing the richest chemistry, similar to W2010, which is common of more passive discs.

Realistically, these comparisons are very weak because of the significantly different dynamics between our model and other models in the literature. In essence, what they show is that in a more passive, axisymmetric $\alpha$-disc, a photon-dominated, chemically rich layer sits above a shielded, cold, chemically sparse midplane. However, in a non-axisymmetric, gravitationally unstable disc, global instabilities and mass transport produce a hot midplane that experiences a rich chemical evolution. Ideally, other 3D simulations of GI-driven discs need to be completed before appropriate comparisons can be made to the results presented in this paper. 

\section[Conclusions \& Future Work]{Conclusion}
\label{sec:Conclusion}

We have modelled the physical and chemical evolution of a $0.17\,\mathrm{M}_{\odot}$ protoplanetary disc over a period of approximately $2000\,\mathrm{yr}$. The disc surrounds a $0.8\,\mathrm{M}_{\odot}$ protostar that will evolve into a Solar-like star. As a result, our work may be indicative of an early dynamical and chemical phase of evolution of our Solar System.

\smallskip

This paper extends previous work that suggested gravitational instabilities significantly affect the chemistry in more massive systems. The main results that we have found from simulating a lower mass, protosolar disc are as follows:

\begin{itemize}[leftmargin = 0.0cm, itemindent = 1.0cm]

\item Spiral waves generated in the disc heat material through shocks that enhance the local temperature and increase the rates of desorption and some endothermic reactions. This subsequently increases the gas-phase abundance of most species across the entire disc, including the midplane.

\item In the more massive disc, the majority of species trace distinct spiral structure in the column density maps after $2.7\,\mathrm{ORPs}$ at $30\,\mathrm{au}$ (defined as $t$ = $t_{orp}$). After the same number of outer rotational periods in the lower mass disc, the majority of species do not trace much spiral structure because the spiral density waves are of lower amplitude.

\item At the end of our simulation spiral features driven by GIs in a protosolar disc are traced by a significant number of important species; CN is a particularly strong tracer as it is most abundant in the outer disc regions and has a relatively high column density.

\item Comparing the less massive and more massive discs at $t$ = $t_{orp}$ reveals that most species have reduced or enhanced peak abundances by a factor of only two to four times.  Therefore, this suggests that the presence of gravitational instabilities in young, massive protoplanetary discs defines the chemical evolution more significantly than the mass differences of the discs.

\item We have found that GIs, due to global transport of disc material, produce a chemistry-rich midplane. This is in contrast to studies of more evolved, axisymmetric discs, that report peak abundances in a heated layer above the cold `gas-phase desert' midplane, caused by photoprocesses and local transport of material. 

\item Rapid succession of shocks can lead to long-lasting changes in the inner disc composition. At small radii, the shock heating quickly increases the temperature beyond the desorption energy for all neutral species. Therefore we recover abundances close to the protostar that are likely higher than in a more passive disc. 

\item In the outer disc, the effect of shock heating is easily distinguishable because of the lower ambient temperature. For most species this effect is short-lived, however, successive shock heating permanently alters the abundances of HNO, CN and NH$_3$ in the outer disc over the duration of our simulation. This has two major implications. Firstly, chemical simulations of more evolved systems should perhaps consider the processing of GIs when setting initial conditions, if indeed GIs are present in the early phases of disc evolution. Secondly, observations of discs could still be used to characterise disc dynamics even if they can not be resolved spatially as shocks cause transient and permanent changes in gas-phase abundances.

\item H$_2$O destroys HCO$^+$ via charge exchange interaction, creating a hole in HCO$^+$ in the inner disc. The extent of H$_2$O is predominantly determined by instability strength, which is characterised by disc mass. Therefore, observations and measurements of the HCO$^+$ inner hole in real systems could be used as an additional method for characterising protoplanetary disc mass.

\end{itemize}

We have not incorporated envelope accretion or outflows into our model in order to focus on the effects of gravitational instabilities on the chemical evolution of protoplanetary discs. This is appropriate as collimated outflows are assumed to have an insignificant effect on disc composition and a surrounding envelope would likely only enhance the molecular abundances in the outermost regions of the disc. \citet{Sakai2014} have found an enhancement of SO at the centrifugal barrier of a Class 0 object that could be produced from a weakly shocked region as envelope material accretes on to the disc whilst exceeding the Keplerian velocity. Our results support this theory as the gas-phase abundances of volatile species are enhanced through shock heating. However, as the authors find no significant SO abundance interior to the centrifugal barrier, the impact from this phenomenon on our results concerning gravitational instabilities is likely negligible.

\smallskip

Recent observational advancements such as ALMA will revolutionise the studies of protoplanetary discs, and open up the possibilities for investigating young, embedded objects. The resolutions attainable suggest that if the $0.17\,\mathrm{M}_{\odot}$ disc was located within the Taurus-Auriga cloud complex, its spiral structure should be distinguishable, which is an exciting prospect for characterising gravitational instabilities in real systems. As a cautionary note, however, it is possible to have well defined spiral structure and not detect it, or for complicated spiral structure to give an incorrect appearance at low resolution. For example, CO appears to be the best species for directly observing spiral structure in real systems, but at low resolution the image could appear as two rings, which would be incorrectly interpreted as a gap. Therefore, in paper II of this series, we will perform radiative transfer calculations in order to assess the detectability and image fidelity of spiral structure in a lower mass disc, focusing on a combination of the species we have found to be encouraging tracers such as CN, HCS and CO. We will then compare these to the results of \citet{Douglas2013} in an attempt to determine the importance of system mass on observability of such features. 

\section[Acknowledgements]{Acknowledgments}
\label{sec:Acknowledgements}

We would like to thank the anonymous referee for constructive comments that have improved the clarity of this manuscript. MGE gratefully acknowledges a studentship from the European Research Council (ERC; project PALs 320620).  JDI gratefully acknowledges funding from the European Union FP7-2011 under grant agreement no. 284405.  ACB's contribution was supported, in part, by The University of British Columbia and the Canada Research Chairs program.  PC and TWH acknowledge the financial support of the European Research Council (ERC; project PALs 320620).

\appendix
\section[DesTemps]{Desorption Temperatures} 
\label{sec:DesTemps}

The rate of desorption from a surface is given by the Polanyi-Wigner relation \citep{Tielens2005},
\begin{equation}
  R_i = \nu_i e^{-E_i / kT_{\mathrm{gr}}}
\end{equation}
where $E_i$ is the adsorption binding energy of species $i$, $\nu_i$ = 1.6x10$^{11}$ $\sqrt{(E_i/k)/(m_i/m_{\mathrm{H}})}$ s$^{-1}$ is the vibrational frequency of the species in the surface potential well, $k$ is Boltzmann's constant, and $m_i$ and $m_H$ are mass of species $i$ and hydrogen, respectively. $T_{gr}$ is the temperature of the grains, which we assume to be in equilibrium with the gas temperature. The freeze-out temperature can be determined by equating the flux of desorbing molecules to the flux of gas molecules adsorbing on to the grain surface, calculated from gas kinetic theory. Thus,
\begin{equation}
  N_{s,i} R_i f_{s,i} = 0.25 n_i v_i
\end{equation}
where $N_{s,i}$ is the number of adsorption sites per cm$^2$, $f_{s,i}$ is the fraction of the surface adsorption
sites that are occupied by species $i$ (see Ilee2011, Equation 9), $n_i$ is the gas-phase number density of species $i$, and $v_i$ is the thermal speed. A sticking probability of unity is assumed. By solving for the dust temperature we derive the desorption temperature for species $i$, \citep{Hollenbach2009},
\begin{equation}
  T_{des} \simeq \frac{E_i}{k}
    \left[ ln \left(
      \frac{4N_{s,i}f_{s,i}\nu_i}{n_i v_i}
    \right) \right]
\label{eq:Des}
\end{equation}
Using equation \ref{eq:Des} we derived the desorption temperatures for the neutral species featured in this paper, which are shown in Table \ref{tab:DesTempsTable}. We have used an average gas-phase number density for each species in order to provide desorption temperatures that are applicable across the whole disc, within a maximum error of 10 per cent. 

\begin{table}
  \centering
  \caption{Desorption temperatures.}
  \begin{tabular}{ccc}
  \hline
   Species  &  Binding Energy [$\mathrm{K}$]   &   $T_{\mathrm{des}}$ [$\mathrm{K}$] \bigstrut \\
 \hline
    CO   &	855  &  26 \bigstrut[t] \\
    CO2  & 	2575  &  74\\
    SO   &	2600  &  72\\
    SO2  &	3405  &  93\\
    NH3  & 	5534  &  151\\
    H2O  & 	5773  &  167\\
    H2CO & 	2050  &  57\\
    H2S  & 	2743  &  76\\
    HCN  & 	2050  &  57\\
    HNC  & 	2050  &  57\\
    OCS  &	2888  &  79\\
    HCO  & 	1600  &  44\\
    HNO  & 	2050  &  57\\
    CS   &	1900  &  53\\
    HCS  & 	2350  &  65\\
    OCN  & 	2400  &  66\\
    O2   &	1200  &  33\\
    CN   &	1600  &  44 \bigstrut[b] \\
  \hline
  \end{tabular}
  \label{tab:DesTempsTable}
\end{table}

\section[HCOp inner hole]{HCO$^+$ inner hole} 
\label{sec:HCOpH2ORadialAbundances}

\begin{figure*}
    \includegraphics[width = 0.75\textwidth]{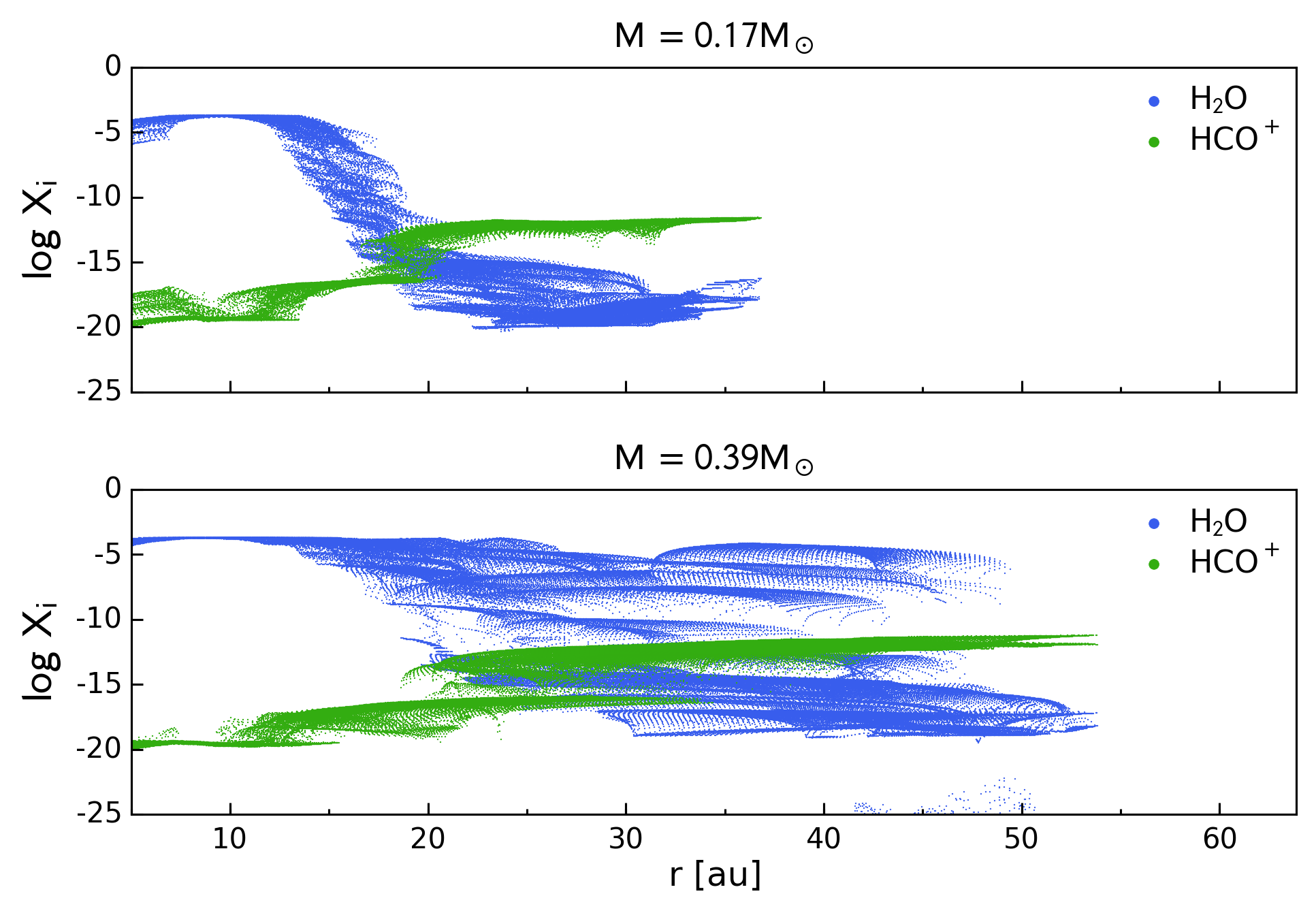}
    \caption{Radial extent of H$_2$O and HCO$^+$ fractional gas-phase abundances in the midplane of the $0.17\,\mathrm{M}_{\odot}$ (top) and $0.39\,\mathrm{M}_{\odot}$ (bottom) discs at $t$ = $t_{\mathrm{orp}}$. The abundances for every midplane cell are plotted, and because of the shock heating and non-axisymmetric nature of the discs, there is a wide range in abundance values at each radii.}
    \label{fig:H2OHCOpAbundances440ILEE}
\end{figure*}

Figure \ref{fig:H2OHCOpAbundances440ILEE} shows the abundances of H$_2$O and HCO$^+$ as a function of midplane radius within the lower mass and higher mass discs at $t$ = $t_{\mathrm{orp}}$. In the more massive disc the overall extent of significantly abundant H$_2$O is much larger due to the higher amplitude spiral waves. As a result, the H$_2$O and HCO$^+$ abundances are enhanced and suppressed, respectively, over a larger fraction of the disc radius, and so the inner hole seen when viewing from above is larger than in the lower mass disc. This result indicates that the measurement of the HCO$^+$ hole size in real systems could potentially be used as a tracer for disc mass.

\section[ShockLHOS]{Removing shocks} 
\label{sec:ShockLHOS}

\begin{figure}
    \includegraphics[width = 0.475\textwidth]{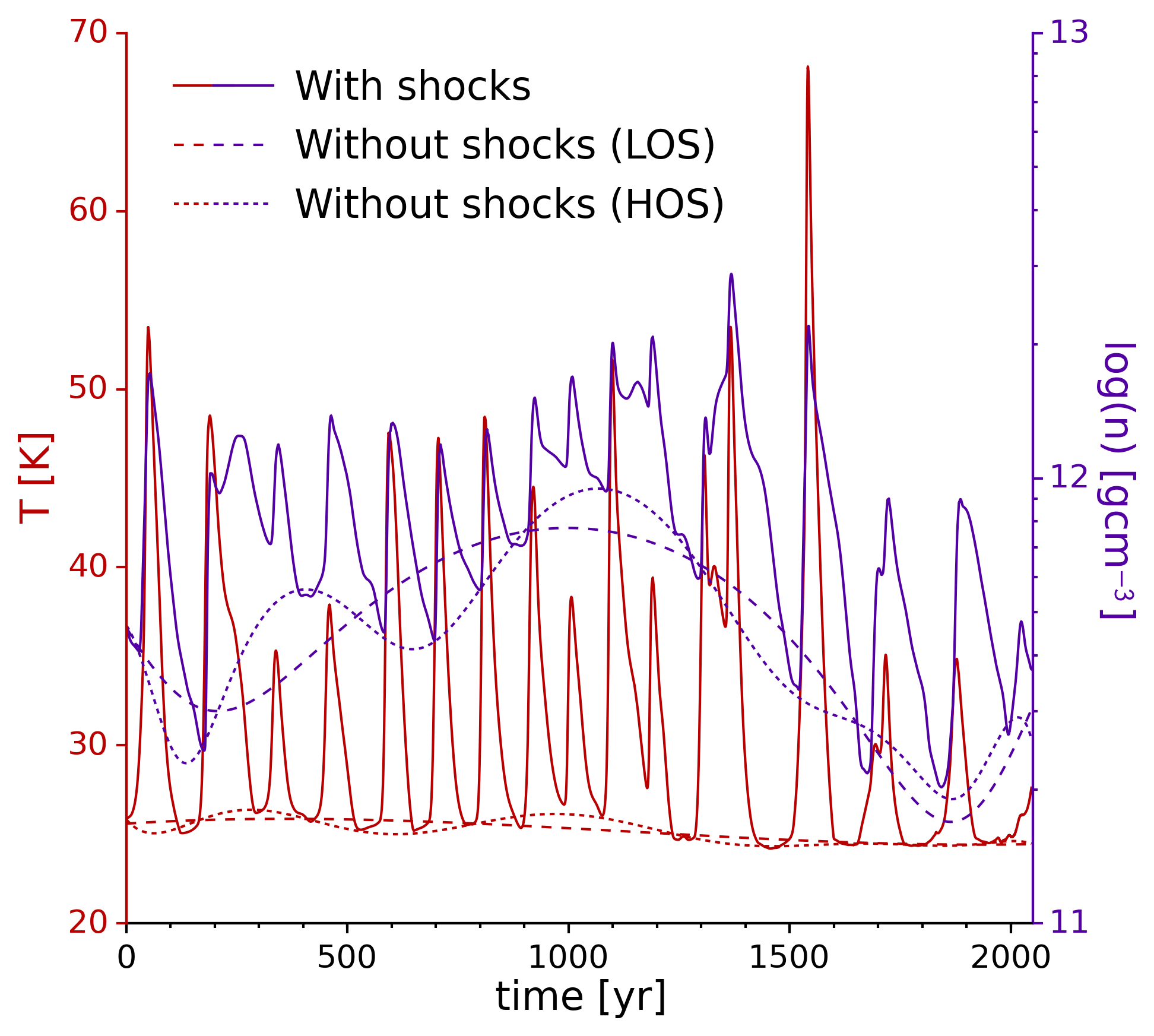}
    \caption{Temperature and number density history of a fluid element in the outer disc. Overplotted are polynomials of low and high-order fitted to the minima of temperature and density, representing the same fluid element history devoid of shocks.}
    \label{fig:ShockLHOSTempDensity}
\end{figure}

Figure \ref{fig:ShockLHOSTempDensity} shows the temperature and number density history of the same fluid element in the outer disc with and without shocks. Artificial fluid elements were constructed by fitting low and high-order polynomials to the minima of temperature and number density, which are more appropriate for trajectories within a passive disc. The abundances of significant species were calculated for the fluid elements with and without shocks, and compared to determine the persistent effects of shock heating within a gravitationally unstable disc.

\end{document}